\documentclass[%
 reprint,
nofootinbib,
 amsmath,amssymb,
 aps,
]{revtex4-1}

\usepackage{hyperref}
\usepackage{todonotes}

\usepackage{graphicx}% Include figure files
\usepackage{dcolumn}% Align table columns on decimal point
\usepackage{bm}% bold math
\usepackage{xcolor}
\newcommand{\M}{\mathcal{M}}
\newcommand{\DA}{\bar{\mathcal{D}}}
\renewcommand{\S}{\mathcal{S}}
\begin{document}

%\preprint{APS/123-QED}

\title{Dark Matter Model \textit{or} Mass, but Not Both:\\
Assessing Near-Future Direct Searches with Benchmark-free Forecasting}

\author{Thomas D. P. Edwards}
\email{t.d.p.edwards@uva.nl}
\author{Bradley J. Kavanagh}
\email{b.j.kavanagh@uva.nl}
\author{Christoph Weniger}
 \email{c.weniger@uva.nl}
\affiliation{%
 Gravitation Astroparticle Physics Amsterdam (GRAPPA), Institute
for Theoretical Physics Amsterdam and Delta Institute for Theoretical Physics,
University of Amsterdam, Science Park 904, 1090 GL Amsterdam, The Netherlands 
}%

\date{\today}% It is always \today, today,
             %  but any date may be explicitly specified

\begin{abstract}
  Forecasting the signal discrimination power of dark matter (DM) searches is
  commonly limited to a set of arbitrary benchmark points.  We introduce new
  methods for benchmark-free forecasting that instead allow an exhaustive
  exploration and visualization of the phenomenological distinctiveness of DM
  models, based on standard hypothesis testing. Using this method, we reassess
  the signal discrimination power of future liquid Xenon and Argon direct DM
  searches.  We quantify the parameter regions where various non-relativistic effective operators,
  millicharged DM, and magnetic dipole DM can be discriminated, and where upper
  limits on the DM mass can be found. We find that including an Argon target
  substantially improves the prospects for reconstructing the DM properties. We also
  show that only in a small region with DM masses in the range
  $20$--$100\,\mathrm{GeV}$ and DM-nucleon cross sections a factor of a
  few below current bounds can near-future Xenon and Argon detectors
  discriminate both the DM-nucleon interaction and the DM mass simultaneously.
  In all other regions only one or the other can be obtained.
\end{abstract}

\pacs{Valid PACS appear here}% PACS, the Physics and Astronomy
                             % Classification Scheme.
%\keywords{Suggested keywords}%Use showkeys class option if keyword
                              %display desired
\maketitle

%\tableofcontents

\textit{Introduction.---} Searching for the elusive Dark Matter (DM) particle
has been the preoccupation of physicists for many years
\cite{Jungman:1995df,Bertone:2004pz}. Over the past decade, two-phase scintillator
direct detection experiments \cite{Goodman:1984dc,Drukier:1986tm} have found
much success with the LUX \cite{Akerib:2016vxi}, XENON \cite{Aprile:2017iyp} and
PANDA-X \cite{Cui:2017nnn} collaborations providing the most stringent
constraints on DM particles in the $\textrm{GeV}-\textrm{TeV}$ mass range to date. Such experiments will continue to improve in
sensitivity for many years to come. In the case of a detection, it should be
possible to study the astro- and particle-physics properties of DM using
a variety of detectors and detection methods (see \cite{Peter:2013aha, Kavanagh:2014rya, Cahill-Rowley:2014boa,
Alves:2015pea, Bertone:2017adx} and many others), but the precise parameter regions in
which these properties can actually be measured is hard to quantify.

Exploring the prospects for discriminating between different DM-nucleon
interactions usually relies on comparing a number of benchmark models
\cite{Catena:2014epa, Gresham:2014vja, Gluscevic:2014vga, Gluscevic:2015sqa,
Kahlhoefer:2016eds, Catena:2017wzu, Baum:2017kfa}.  However, the pair-wise
comparison of different benchmark points in the model parameter space (DM couplings or masses) is time-consuming, does not scale well with the number
of benchmark points, and is in particular problematic in high-dimensional
parameter spaces. In direct detection, such a high-dimensional parameter space
appears in the framework of non-relativistic effective field theory (NREFT)
\cite{Fan:2010gt,Fitzpatrick:2012ix,Fitzpatrick:2012ib,Anand:2013yka,Dent:2015zpa},
in which the space of DM-nucleon interactions may have more than 30 dimensions \cite{Catena:2014hla, Catena:2014uqa,
Catena:2015uua}.  With current techniques, it is hence difficult to study model
degeneracies and the degeneracy-breaking power of future instruments in a
reliable and exhaustive way.  For such tasks, dedicated techniques are
required.

\smallskip

In this \textit{Letter}, we introduce a new framework for studying the signal
discrimination power of future detectors in a fundamentally benchmark-free way.
The key questions we aim to address are:  How many \textit{observationally
distinct} signals does a given model predict for a set of future experiments?
How many of these signals are compatible with specific subsets of the signal
model?  In which regions of parameter space is signal
discrimination and parameter reconstruction possible?

We first summarize the basics of our approach.  We then discuss the dark matter models and experiments we consider in the current work. Finally, we show our results 
and conclude with a short discussion about possible future directions and applications\footnote{Code associated with the paper available at \url{https://github.com/tedwards2412/benchmark_free_forecasting/}.}.

\smallskip

\textit{Information Geometry.---}
Consider a New-Physics model $\M$ with some $d$-dim model parameter space,
$\vec\theta \in \Omega_\M\in\mathbb{R}^d$, and a combination of future
experiments $X$ that are described by some likelihood function
$\mathcal{L}_X(\mathcal{D}|\vec\theta)$, where $\mathcal{D}$ is data.  We
expect that two model parameter points $\vec\theta, \,\vec\theta'$  can be
discriminated by experiments $X$ if the parameter point $\vec\theta'$ is
inconsistent with Asimov data \cite{Cowan:2010js} $\mathcal{D} = \DA(\vec\theta)$.  More
concretely, distinctiveness requires that the log-likelihood ratio
\begin{equation}
  \text{TS}(\vec \theta')_{\DA(\vec\theta)} \equiv- 2 \ln
  \frac{\mathcal{L}(\DA(\vec\theta)|\vec \theta')}{
  {\displaystyle \max_{\vec\theta''}}\mathcal{L}(\DA(\vec\theta)|\vec\theta'')}
  \simeq \lVert \vec x(\vec\theta) - \vec x(\vec\theta') \rVert^2\,,
  \label{eqn:TS}
\end{equation}
exceeds a threshold value $r_{\alpha}(\M)^2$. The threshold value depends on the chosen
statistical significance, which we set here to $\alpha = 0.045$ ($2\sigma$), as
well as the sampling distribution of $\text{TS}(\vec\theta')$.  In the
large-sample limit and under certain regularity conditions, the sampling
distribution follows a $\chi^2_k$ distribution with $k=d$ degrees of
freedom~\cite{Wilks:1938dza,Cowan:2010js}.

The last part of Eq.~\eqref{eqn:TS} is an approximation based on the
`euclideanized signal' method \cite{Edwards:2017kqw}, an embedding $\vec\theta \mapsto \vec
x(\vec\theta) \in \mathbb{R}^n$ into some, usually higher-dimensional, space
with unit Fisher information matrix ($n$ usually equals the total number of
data bins).  This approximation maps statistical distinctiveness onto euclidean
distances, and works to within 20\% if the number of counts is order one, see \cite{Edwards:2017kqw} for a discussion and caveats.  Confidence regions in the model parameter space correspond then to
hyperspheres of radius $r_{\alpha}(\M)$ in the euclideanized signal space.

\begin{figure}[t]
  \includegraphics[width=.5\textwidth]{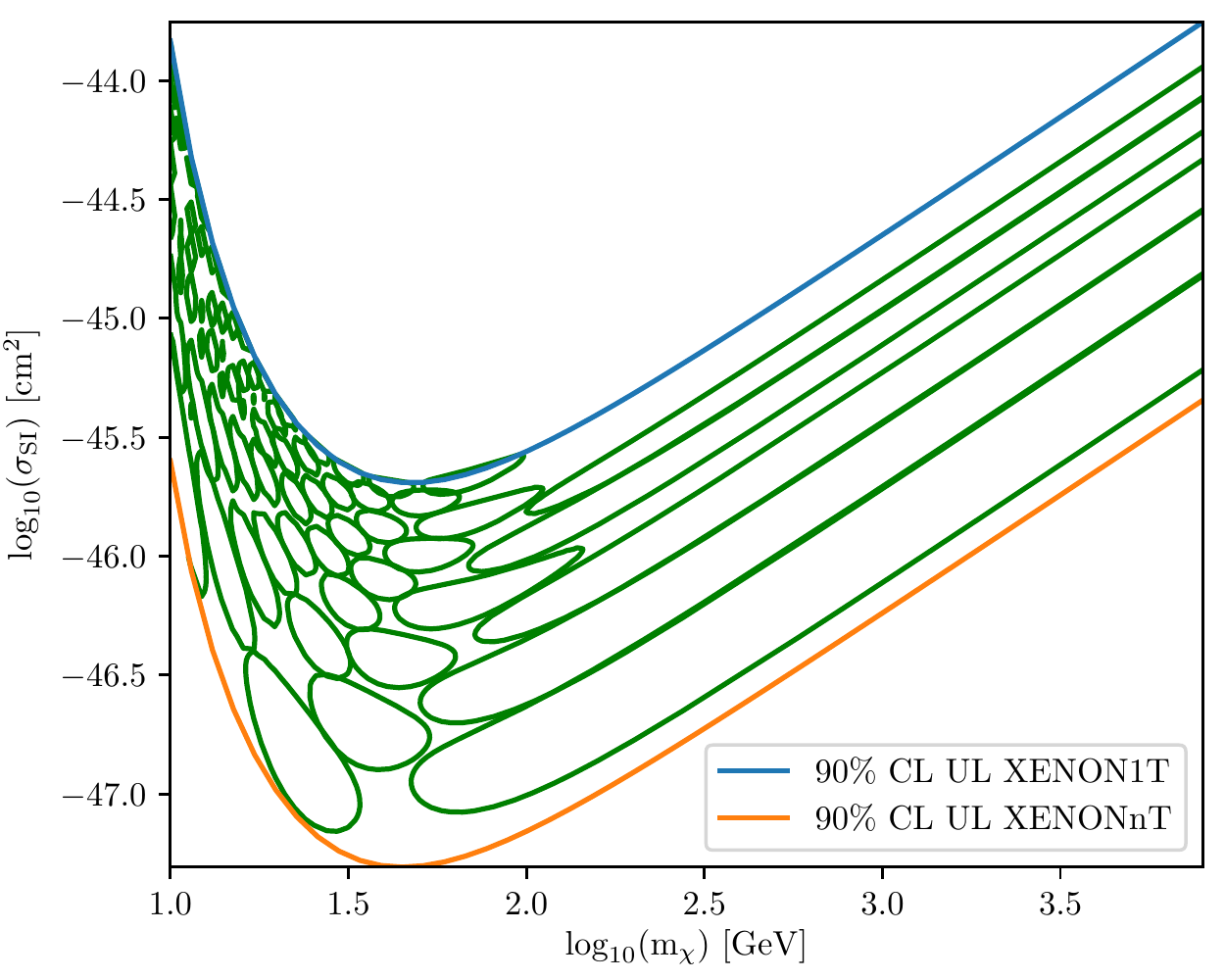}
  \caption{Limit on the SI cross-section vs.~DM mass, assuming operator $\mathcal{O}_1$, for a XENONnT detector. Contours show $ 68\%$ confidence contours in $d=2$ dimensions. The radius of these contours in the Euclidean space is therefore $r_{\alpha}(\M) = 1.52$. The number of discriminable signals in the blue+blue/orange region of the middle panel of Fig.~\ref{fig:Venn} can be approximated by counting the number of closed green contours.}
\label{fig:regions}
\end{figure}

\smallskip

Often one is interested in sub-models $\S$ that are nested inside model $\M$, and
which are obtained by restricting $\M$ to a $d'$-dim subregion $\Omega_{\S}
\subset \Omega_\M$.  A parameter point $\vec\theta$ of $\M$ leads to a
signal that is distinct from any signal in submodel $\S$ if
\begin{equation}
  - 2 \ln \frac{\max_{\vec\theta'\in \Omega_{\S}}\mathcal{L}
  (\DA(\vec\theta)|\vec \theta')}{
  \max_{\vec\theta'}\mathcal{L}(\DA(\vec\theta)|\vec\theta')}
  \simeq
  \min_{\vec\theta'\in \Omega_{\S}} 
  \lVert \vec x (\vec\theta) - \vec x(\vec\theta')\rVert^2\,,
  \label{eqn:pTS}
\end{equation}
exceeds a certain threshold value $r_{\alpha}(\S,\M)^2$.  Here `distinct' means
that the composite null hypothesis $\S$ can be rejected for data $\DA(\vec\theta)$.
The sampling distribution of Eq.~\eqref{eqn:pTS} follows in general a $\chi^2_{k = d-d'}$
distribution.  In the euclideanized signal space $\vec x$, parts of model
$\M$ cannot be discriminated from model $\S$ that lie within a `shell' of thickness
$r_{\alpha}(\S,\M)$ around the signal manifold of $\S$.

Finally, nuisance parameters can be accounted for by replacing the likelihood
function in Eq.~\eqref{eqn:TS} with a profile likelihood,
$\mathcal{L}(\mathcal{D}|\vec\theta)=
\max_{\vec\eta}\mathcal{L}(\mathcal{D}|\vec\theta,
\vec\eta)\mathcal{L}_\eta(\vec\eta)$, where the last factor can incorporate
additional constraints on the nuisance parameters from data external to $X$.

\smallskip

\textit{Distinct signals.---}
To quantify the signal discrimination power of a set of future experiments $X$
in the context of model $\M$ we may define the figure of merit
\begin{equation}
  \nonumber
  \nu^\alpha_{\mathcal{M},X}(\Omega_\M) \simeq \parbox{6.0cm}{
    Total number of signals from model $\M$ discriminable by 
    experiments $X$.}
  \label{eqn:V}
\end{equation}
More specifically, $\nu^\alpha_{\M,X}$ equals the maximum number of points that
can populate the parameter space of $\M$ while remaining mutually
distinct according to Eq.~\eqref{eqn:TS}.
Any such set of points provides a complete sample of the phenomenological features of model $\M$.  Loosely speaking, the
points correspond to a set of non-overlapping confidence contours as shown in
Fig.~\ref{fig:regions}.  
Furthermore, when considering a sub-model $\S$ nested in $\M$, we can define
\begin{equation}
  \nu^\alpha_{\M, X}(\Omega_\S) = \parbox{6cm}{
    Number of signals from model $\M$ discriminable by 
    experiments $X$, and consistent with model $\S$.}
  \label{eqn:Vsub}
\end{equation}
With these definitions, the phenomenological distinctiveness of various regions
in the parameter space of model $\M$ can be visualized using standard Venn
diagrams \cite{Venn1880}.  The
technical definition for the measure $\nu^\alpha_{\M, X}(\cdot)$, which is used
in the subsequent examples, is given in Appendix~A of the Supplemental Material.

\begin{figure*}[t]
    \includegraphics[width=.32\textwidth]{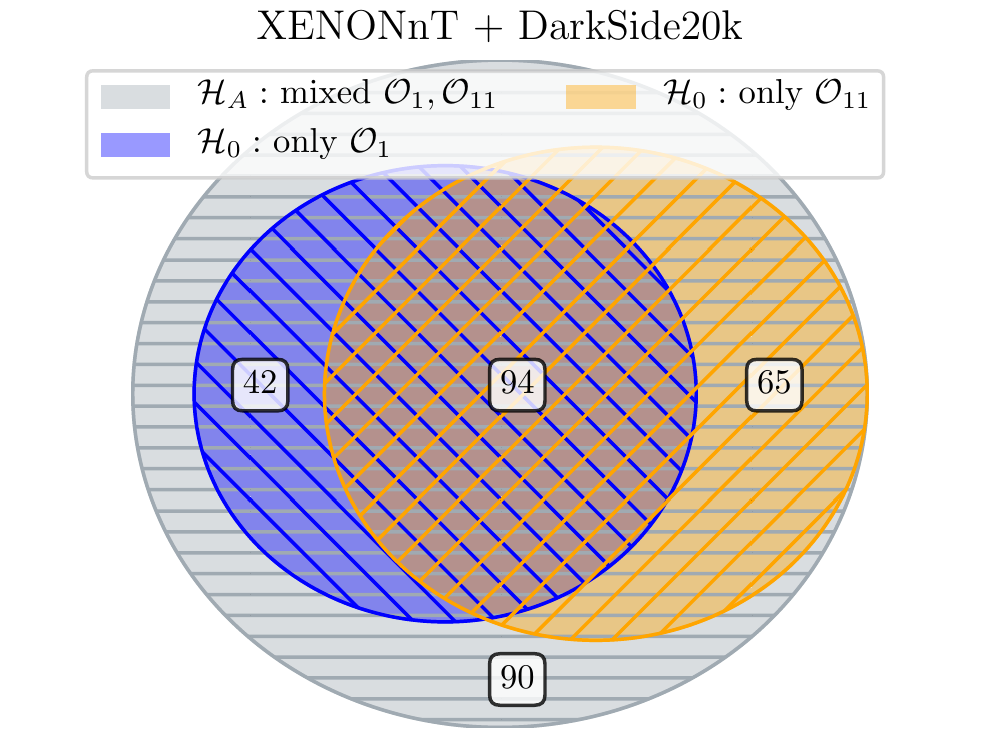}
  \includegraphics[width=.32\textwidth]{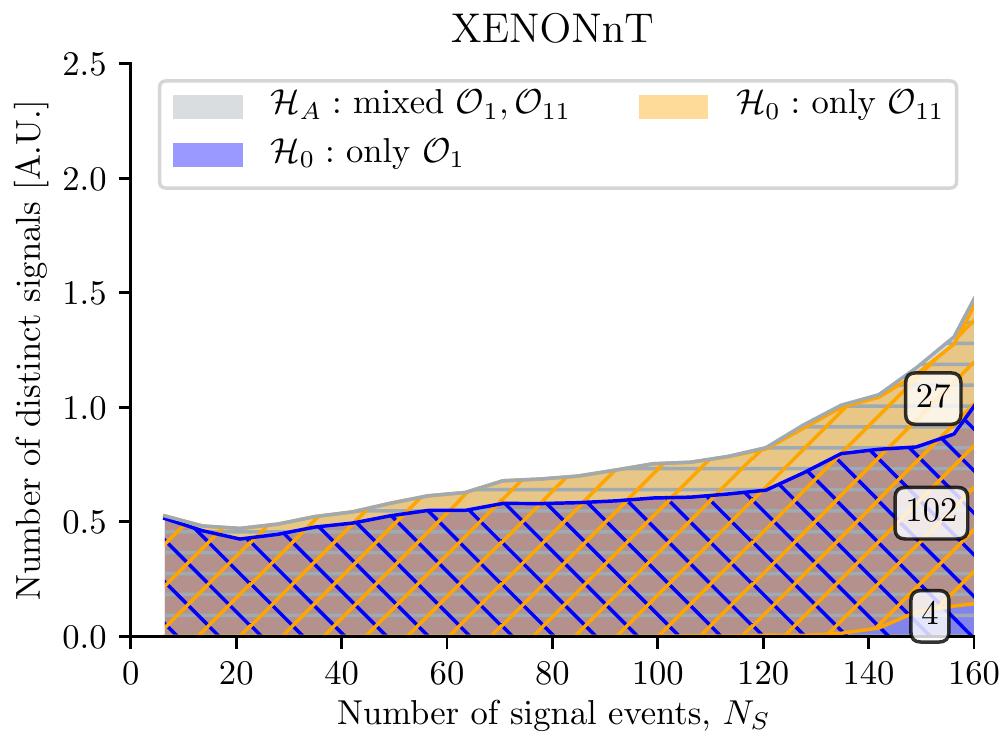}
  \includegraphics[width=.32\textwidth]{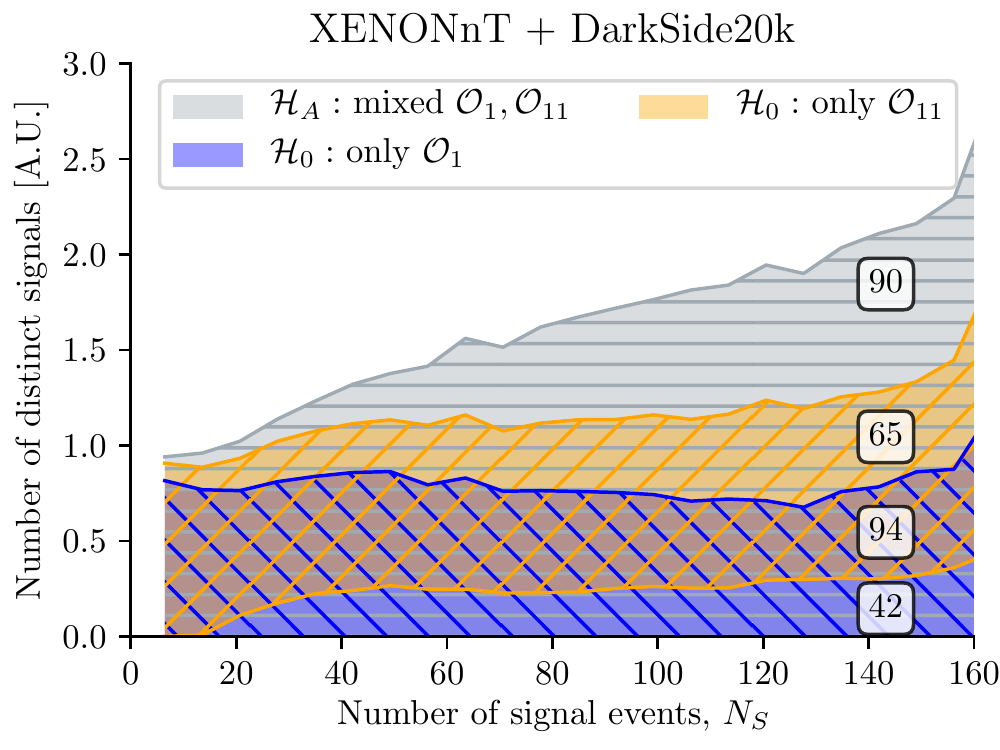}
  \caption{Signal-discrimination power of combinations of direct detection experiments, summarized in \textit{infometric Venn diagrams}. \textit{Central/Right panel:} The full area corresponds to the \textit{number of distinct detectable signals} within the alternative hypothesis $\mathcal{H}_A$, unfolded as function of number of XENONnT signal events.  The subsets indicate the fraction of signals consistent with various null hypotheses $\mathcal{H}_0$.  
Overlapping subsets correspond to signals consistent with multiple null hypotheses simultaneously. The numbers correspond to $\nu_{\M, X}(\Omega_\M)$ in each region.  In the right panel, the overlapping (blue+orange) region corresponds to the model parameters between the purple-dot-dashed and orange contours in Fig.~\ref{fig:comp}. The non-overlapping (blue-only) $\mathcal{O}_1$ region corresponds to parameters between the purple-dashed and blue contours in Fig.~\ref{fig:comp}. \textit{Left panel:} Standard Venn diagram summed over number of signal events.
}
\label{fig:Venn}
\end{figure*}

\begin{figure}[t]

\includegraphics[width=0.5\textwidth]{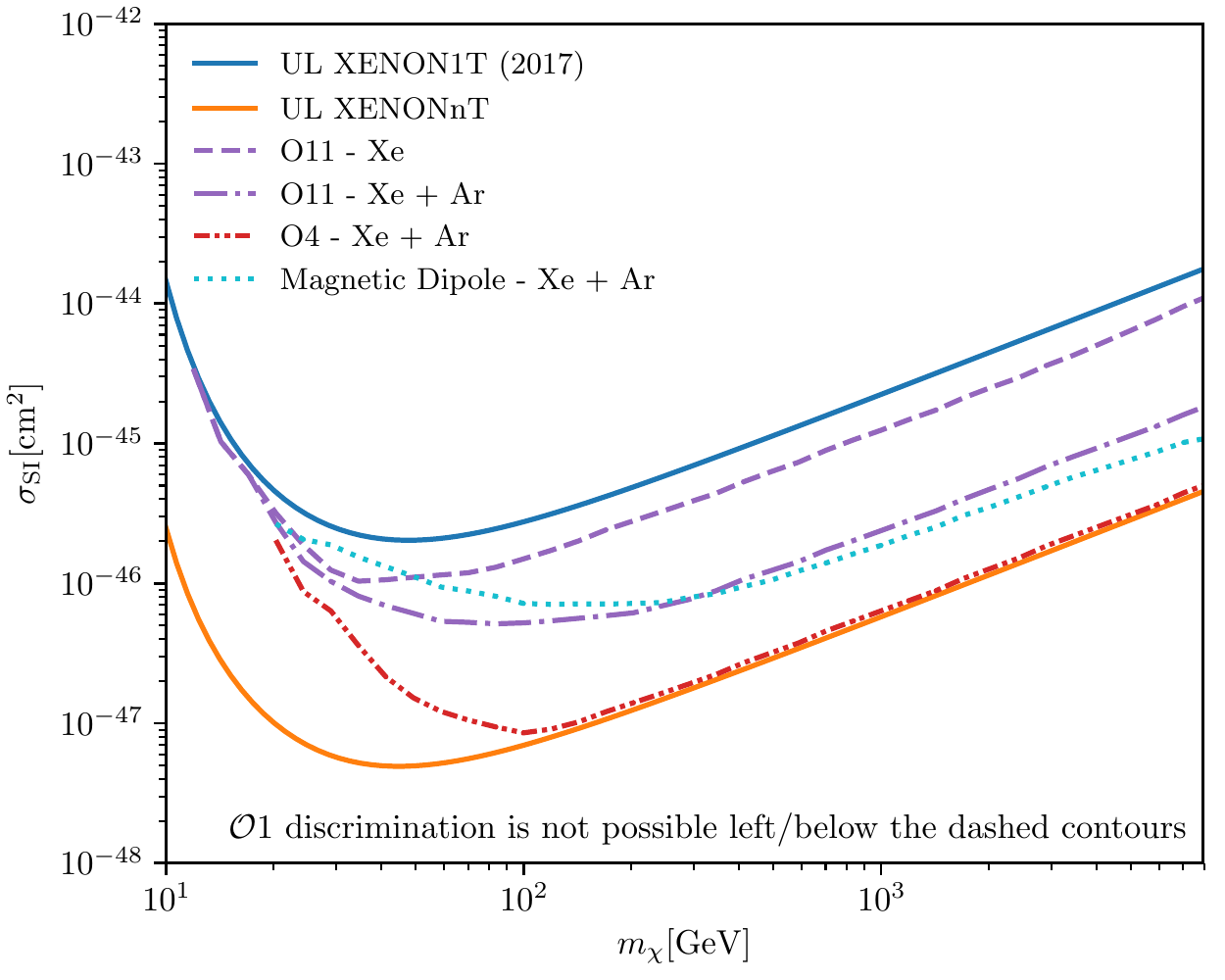}
\caption{Discriminability of DM interactions. To the left/below each broken line, it is not possible to discriminate an $\mathcal{O}_1$-signal (with the indicated cross-section and DM mass) from the corresponding best-fit $\mathcal{O}_4$, $\mathcal{O}_{11}$ or magnetic dipole signal.  Above/right of each broken line, such a discrimination is possible with at least $2\sigma$ significance.   
Solid lines display 90\% CL limits for XENON1T-2017 and XENONnT. 
All lines include DM halo uncertainties.}
\label{fig:comp}
\end{figure}

\begin{figure}[t]
\includegraphics[width=.5\textwidth]{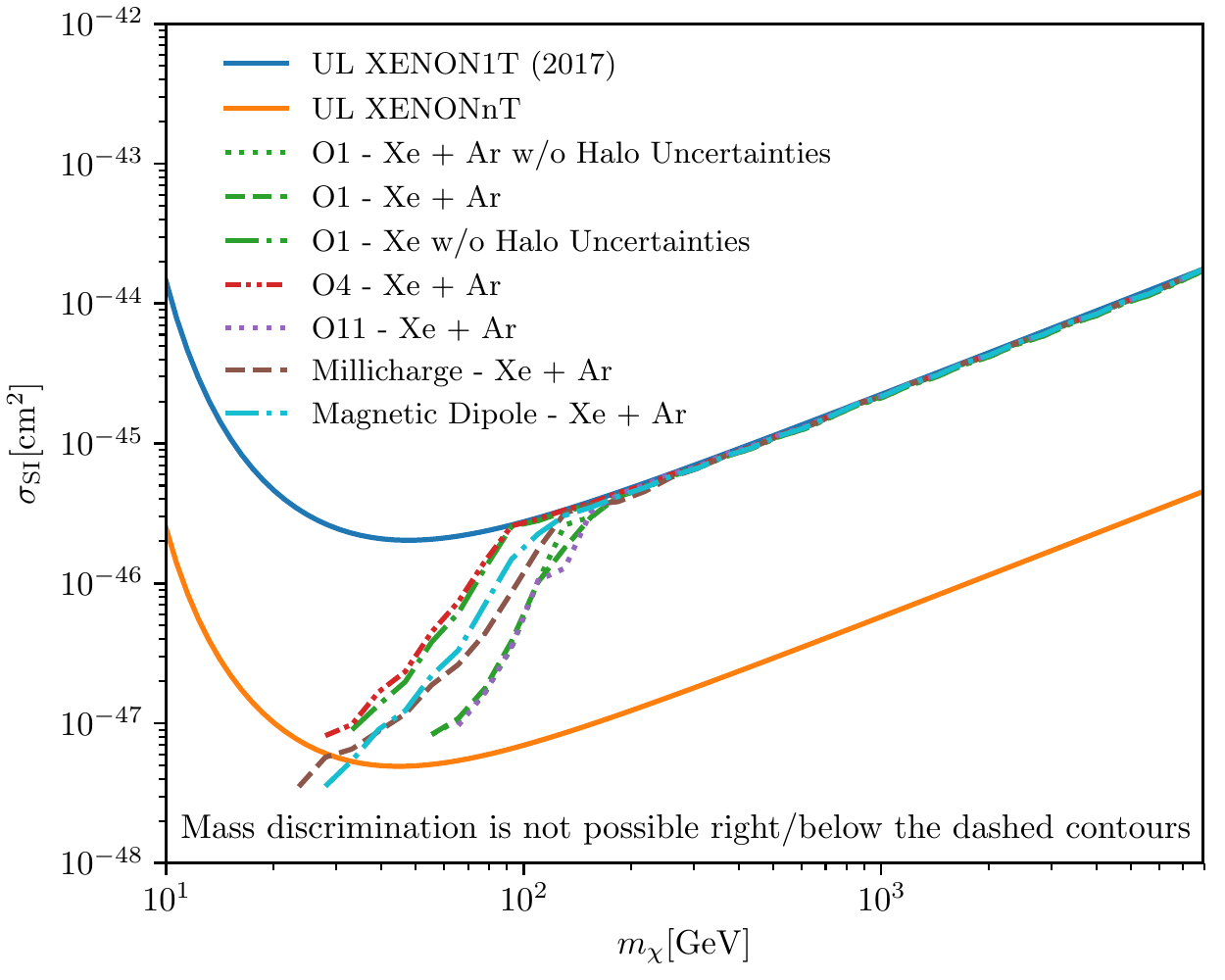}
\caption{Discriminability of the DM mass. To the right of each broken line, the DM mass is unbounded from above (at
  $2\sigma$).  
Lines for other operators are mapped onto $\sigma_{\mathrm{SI}}$ by
converting to an effective cross section and rescaled to
match $\mathcal{O}_1$ at high masses.  For $\mathcal{O}_1$, we
also show constraints for the
  Xenon-only case and when halo uncertainties are neglected.  Solid lines display 90\% CL limits for XENON1T-2017 and XENONnT.}
\label{fig:mass}
\end{figure}

\smallskip

\textit{DM-nucleon interactions.---}
While direct detection is typically analysed in terms of the standard
spin-dependent (SD) and spin-independent (SI) interactions
\cite{Cerdeno:2010jj}, the range of possible signals is much broader. The
framework of non-relativistic effective field theory (NREFT) \cite{Fan:2010gt,Fitzpatrick:2012ix,Fitzpatrick:2012ib,Anand:2013yka,Dent:2015zpa} aims to classify possible elastic DM-nucleon interactions and thus possible
signals in DM-nucleus scattering experiments. NREFT is realised as a power
series in the DM-nucleus relative velocity $\vec{v}$ and the nuclear recoil
momentum $\vec{q}$, valid for non-relativistic, short-range
interactions. The resulting operators (labelled $\mathcal{O}_1,\,
\mathcal{O}_3, \, \mathcal{O}_4,\,...$) give rise to a range of novel energy
spectra \cite{Chang:2009yt,Gresham:2014vja,Gluscevic:2015sqa,Baum:2017kfa},
directional signals \cite{Catena:2015vpa,Kavanagh:2015jma} and annual
modulation signatures \cite{DelNobile:2015tza,DelNobile:2015rmp}. We focus here on the three operators  $\mathcal{O}_1$,
$\mathcal{O}_4$ and $\mathcal{O}_{11}$ because they allow us to explore a
diverse range of signals with only a small number of operators\footnote{We neglect the effects of operator mixing \cite{DEramo:2016gos,Bishara:2017nnn} which require us to specify the structure of the dark sector.}. Operator
$\mathcal{O}_1 = 1_\chi 1_N$ couples to nucleon number while the operator
$\mathcal{O}_4 = \vec{S}_\chi \cdot \vec{S}_N$ couples to nuclear spin,
allowing us to explore the complementarity between nuclei of different size and
spin \cite{Cerdeno:2013gqa}. Operator $\mathcal{O}_{11} = i \vec{q}\cdot
\vec{S}_\chi /m_N$ may arise as the leading-order interaction in certain
scalar-mediated models \cite{Dent:2015zpa}. Similar to $\mathcal{O}_1$, it also
couples to nucleon number and receives a coherent enhancement to the rate, but
has a characteristic peaked recoil spectrum owing to an extra
$\mathrm{d}\sigma/\mathrm{d}E_R \propto E_R$ scaling of the cross section
\cite{Gluscevic:2015sqa}. This allows us to explore how well different recoil
spectra can be discriminated in future experiments.

Unfortunately, NREFT cannot encompass all possible
signals. In particular, in its original formulation \cite{Fitzpatrick:2012ix} it cannot describe interactions through light mediators. In this case, the
typical momentum transfer is larger than the mediator mass and an expansion in
$q$ is no longer appropriate\footnote{Note, however, that because the DM is still non-relativistic, the effects of light mediators can be incorporated into the NREFT by including the appropriate propagator \cite{DelNobile:2013sia,Kahlhoefer:2017ddj}.}. 
The scenario in which this mediator is the
Standard Model photon has been studied extensively \cite{Ho:2012bg,
DelNobile:2013sia, DelNobile:2014eta}. Here, we consider millicharged DM \cite{McDermott:2010pa} which has long-ranged,
coherently-enhanced interactions with nuclei, with a differential cross section
scaling as $E_R^{-2}$ \cite{Fornengo:2011sz, DelNobile:2013sia, Panci:2014gga}.
Alternatively, DM may have non-zero electric and magnetic moments
\cite{Sigurdson:2004zp, Banks:2010eh}, particularly if it takes the form of a
composite state, such as a Dark Baryon \cite{Bagnasco:1993st, Banks:2005hc}. In
the context of model discrimination, most interesting for us
will be the magnetic dipole, $(\mu_\chi/2)\, \overline{\chi}
\sigma^{\mu \nu} \chi \, F_{\mu \nu}$, which leads to both long-range and
short-range contributions to the rate, arising from charge-dipole and
dipole-dipole interactions respectively \cite{Barger:2010gv, DelNobile:2012tx,
DelNobile:2014eta}.

\smallskip

The five DM-nucleon interaction models we have outlined above encompass a range
of phenomenologically-driven as well as more theoretically-motivated models, leading to a wide range of direct detection signals. We calculate the signal
spectra in each case using the publicly-available code
\textsf{WIMpy}
\cite{WIMpy-code}, implementing expressions from \cite{Anand:2013yka} and \cite{DelNobile:2013sia}\footnote{Note that
the operator normalisations in \cite{DelNobile:2013sia} and \cite{Anand:2013yka} differ.
}. The required
nuclear response functions are taken from the mathematica package provided in \cite{Anand:2013yka}, supplemented by those calculated in \cite{Catena:2015uha}. We assume iso-spin conserving ($c_p = c_n$) NREFT
interactions and that the particle producing a signature makes up 100\% of the local DM density (which we fix to $\rho_\chi = 0.3\,\,\mathrm{GeV}\,\mathrm{cm}^{-3}$ \cite{Read:2014qva,Green:2017odb}). We incorporate standard Gaussian halo uncertainties from \cite{Green:2017odb}; details can be found in Appendix~B of the Supplemental Material.

% We incorporate halo uncertainties \cite{Green:2017odb} by assuming Gaussian
% likelihood distributions for three parameters of the Maxwellian velocity
% distribution of DM: the Sun's speed $v_\odot= (242 \pm 10)\, \mathrm{km/s}$
% \cite{Bovy:2012ba}, the local circular speed $v_\mathrm{c} = (220 \pm 18) \,
% \mathrm{km/s}$  \cite{Koposov:2009hn}, and the Galactic escape speed
% $v_\mathrm{esc} =(533 \pm 54)\, \mathrm{km/s}$ \cite{Piffl:2013mla}. We assume
% that these uncertainties are uncorrelated \cite{Peter:2009ak}, though in general correlations coming from the
% modeling of the Milky Way halo can be included
% \cite{Strigari:2009zb,Pato:2012fw}. We sample from these distributions as nuisance parameters
% in our signal calculation and include an additional penalization term to the euclideanized signal in~\eqref{eqn:TS} (see Appendix.~\ref{sec:apx}).

\smallskip

\textit{Direct dark matter searches.---}
We implement two toy detectors, designed to resemble the expected
advancement in direct DM searches over the upcoming 5-10 years.

We implement a Xenon detector in light of the stringent constraints on $\mathcal{O}(10) \,\,\mathrm{GeV}$ DM set by XENON1T \cite{Aprile:2017iyp}, with numerous Xenon-based experiments on the horizon \cite{Akerib:2015cja,Aalbers:2016jon,XENONnT,Akerib:2018lyp}. We model this detector on the future \textit{XENONnT} \cite{XENONnT} experiment.
In addition we implement a
detector containing a target material with no nuclear spin, namely Argon, modeling this detector on \textit{Darkside20k} \cite{Aalseth:2017fik}. In this way we
maximize discriminability of spin-dependent operators\footnote{Liquid noble detectors typically do not have sensitivity to DM particles lighter than a few GeV, so we restrict our attention to $m_\chi > 10 \,\,\mathrm{GeV}$ in the current work.}. Our detector implementations and background assumptions are
briefly described in Appendix~B of the Supplemental Material.

\smallskip

\textit{Results.---}  In Fig.~\ref{fig:regions}, we show the expected $68\%\,$CL reconstruction regions for a set of mutually distinct parameter points, for our \textit{XENONnT}-like detector.  The confidence regions are constructed by querying spheres with radius $r_\alpha(\M) = 1.52$ in the euclideanized signal space.  The number of these regions corresponds approximately to the figure of merit in Eq.~\eqref{eqn:V}.  

In the central panel of Fig.~\ref{fig:Venn}, we illustrate the power of \textit{XENONnT} to discriminate between operators in the 3-dimensional model space $\M$ of mass, $\mathcal{O}_{1}$, and $\mathcal{O}_{11}$. With increasing numbers of events, the number of discriminable signals increases, though the majority of signals are compatible with both $\mathcal{O}_{1}$ and $\mathcal{O}_{11}$. In the right panel of Fig.~\ref{fig:Venn}, we include also information from \textit{DarkSide20k}. 
The addition of an Argon detector not only increases the number of discriminable signals (from 133 to 291) but also enlarges the region of parameter space where $\mathcal{O}_{1}$ and $\mathcal{O}_{11}$ can be discriminated from each other. The left panel of Fig.~\ref{fig:Venn} corresponds to the same scenario as the right panel, instead summed over the number of \textit{XENONnT} signal events. $160$ signal events approximately corresponds the expected number of events in XENON-nT if the true model were at the current sensitivity. We note that the Venn diagrams we have introduced here significantly increase in complexity when comparing a large number of models at once. However, we emphasize that the number of discriminable regions, Eq.~\eqref{eqn:V}, is completely general and remains a useful measure for assessing model discriminability.
%become problematic for high-dimensional models. However, we emphasize that the number of discriminable regions (Eq.~\eqref{eqn:V}) is completely general and remains a useful measure even for complicated high-dimensional models.

Figure~\ref{fig:comp} shows the regions of the parameter space of spin-independent ($\mathcal{O}_1$) DM in which discrimination from $\mathcal{O}_4$, $\mathcal{O}_{11}$ and magnetic dipole DM would be possible.  For $\mathcal{O}_4$ (spin-dependent), $2\sigma$ discrimination is possible  at high DM mass even down to small cross sections, when both Xenon and Argon experiments are used. The spin-zero Argon nucleus has no spin-dependent coupling, so we can discriminate well as long as the Argon detector has sensitivity ($m_\chi > \mathcal{O}(20 \,\,\mathrm{GeV})$, below which most recoils are below the 32 keV threshold).

For $\mathcal{O}_{11}$, discrimination is possible at high mass because of the different spectral shapes of $\mathcal{O}_1$ and $\mathcal{O}_{11}$, though cross sections around $\sim 10^{-46}\,\,\mathrm{cm}^2$ are required to obtain enough events to map out the spectra precisely. At low mass, the peak in the $\mathcal{O}_{11}$ spectrum falls below the threshold of the experiments; for both $\mathcal{O}_{1}$ and $\mathcal{O}_{11}$ the exponentially falling tail of the DM velocity distribution dominates the spectral shape \cite{Lewin:1995rx}, making discrimination impossible.

For Magnetic Dipole interactions, discrimination is also possible at high mass, given enough signal events. We note a `kink' in the boundary for Magnetic Dipole interactions around $m_\chi \sim 20 \,\,\mathrm{GeV}$. For large DM masses, the short-range spin-dependent dipole-dipole contribution begins to dominate~\cite{DelNobile:2014eta}. In this case, discrimination prospects are good with the inclusion of the spin-zero Argon detector. 

For the mock detectors we consider, SI interactions cannot be distinguished from Millicharged DM, which is not shown in Fig.~\ref{fig:comp}. The recoil spectrum for Millicharged DM is similar to $\mathcal{O}_1$, but has an extra $E_R^{-2}$ suppression. This more rapidly falling recoil spectrum can be mimicked by an SI interaction with smaller DM mass. As demonstrated in Refs.~\cite{Gluscevic:2015sqa,Kahlhoefer:2017ddj}, low-threshold semi-conductor experiments are required to distinguish between the two interactions.

Finally, Fig.~\ref{fig:mass} shows, for various operators, the regions of parameter space where a closed contour for the DM mass would be possible at the $2\sigma$ level.
At large DM masses, the kinematics of the interaction mean that the recoil spectrum becomes independent of the DM mass, meaning that to the right of the curves in Fig.~\ref{fig:mass}, it is not possible to obtain an upper limit on $m_\chi$ \cite{Green:2007rb,Green:2008rd}. For $\mathcal{O}_1$ we show the regions for Xenon-only, as well as for Xenon and Argon combined without halo uncertainties to demonstrate the improvement in mass reconstruction when including a second detector. When the two detectors are combined halo uncertainties make little difference to the mass discrimination, as changes in the velocity distribution affect the spectra in the two detectors differently \cite{Peter:2013aha}.  
Operator $\mathcal{O}_4$ contains the largest region in which the mass cannot be constrained due to the lack of signal in Argon. 

Even in the most optimistic case of cross sections just below the current bounds, it is not possible to pin down the DM mass for $m_\chi \gtrsim 100\,\,\mathrm{GeV}$. Previous works have demonstrated, typically using a small number of benchmarks \cite{Green:2008rd,Newstead:2013pea,Gluscevic:2015sqa}, that DM mass reconstruction worsens for large masses; here, we have mapped out precisely where this mass reconstruction fails as a function of mass and cross section.

\smallskip

\textit{Discussion.---} The 
methods introduced in
this paper allow us to efficiently characterize and visualize the
phenomenological distinctiveness of direct DM signal models in
infometric Venn diagrams, as shown in Fig.~\ref{fig:Venn}. Furthermore, these
methods allow for an efficient exploration of the phenomenology of
complex models, and hence allow us to make `benchmark-free' statements like
those shown in Figs.~\ref{fig:comp} and~\ref{fig:mass}. 
In Fig.~\ref{fig:comp} we see that ruling out non-standard interactions is harder
for light DM, while in Fig.~\ref{fig:mass} we see that we cannot pin
down the DM mass for masses larger than $\sim 100 \,\,\mathrm{GeV}$
This leaves only a small region of parameter space -- for $m_\chi \in [20,
100]\,\,\mathrm{GeV}$ and cross sections a factor of a few below current bounds
-- in which the DM mass and interaction can both be constrained at the $2\sigma$
level with near-future Xenon and Argon detectors. Such general statements would
not have been possible without an efficient exploration of the Dark Matter
parameter space, made feasible with the tools presented here. Third generation experiments such as DARWIN \cite{Aalbers:2016jon} will have a far greater sensitivity. More events would dramatically improve our ability to constrain different models, particularly for models at the current XENON1T bound.

This work paves the way for a more complete exploration of the direct detection parameter space and a deeper understanding of the complementarity between detectors. Future work should explore the possibility to discriminate between a wider range of interactions, beyond the subset of five we include here. In addition, the techniques we present may be used to optimize detector properties (target material, thresholds, etc.) in order to understand how operator discrimination can be improved at low DM mass. 

Our `benchmark-free' method rests on the 'euclideanised signal' introduced in \cite{Edwards:2017kqw}, and works for any Poisson (and hence Gaussian) likelihood function, as long as background uncertainties are sufficiently Gaussian. Euclideanised signals therefore provide a useful forecasting tool for a wide range of New-Physics signals, including those in cosmology, indirect DM detection, and collider searches. As we have shown, using direct detection alone may not allow us to completely constrain the DM properties. Combining complementary information from other search strategies, coupled with new techniques for efficient forecasting, will provide essential guidance in the future of Dark Matter detection.

\acknowledgments{
The authors thank Riccardo Catena for sharing fortran routines to calculate nuclear structure functions. In addition we thank the GAMBIT collaboration for extensive discussions on future direct detection instruments. We also thank Felix Kahlhoefer, Christopher McCabe, and Mauro Valli for very helpful comments on this manuscript. Finally, we thank the python scientific computing packages numpy \cite{Oliphant:2015:GN:2886196} and scikit-learn \cite{scikit-learn}. This research is funded by NWO through the VIDI research program "Probing the Genesis of Dark Matter" (680-47-532; TE, CW, BK).}

\bibliography{signal-diversity}

%merlin.mbs apsrev4-1.bst 2010-07-25 4.21a (PWD, AO, DPC) hacked
%Control: key (0)
%Control: author (8) initials jnrlst
%Control: editor formatted (1) identically to author
%Control: production of article title (-1) disabled
%Control: page (0) single
%Control: year (1) truncated
%Control: production of eprint (0) enabled
\begin{thebibliography}{78}%
\makeatletter
\providecommand \@ifxundefined [1]{%
 \@ifx{#1\undefined}
}%
\providecommand \@ifnum [1]{%
 \ifnum #1\expandafter \@firstoftwo
 \else \expandafter \@secondoftwo
 \fi
}%
\providecommand \@ifx [1]{%
 \ifx #1\expandafter \@firstoftwo
 \else \expandafter \@secondoftwo
 \fi
}%
\providecommand \natexlab [1]{#1}%
\providecommand \enquote  [1]{``#1''}%
\providecommand \bibnamefont  [1]{#1}%
\providecommand \bibfnamefont [1]{#1}%
\providecommand \citenamefont [1]{#1}%
\providecommand \href@noop [0]{\@secondoftwo}%
\providecommand \href [0]{\begingroup \@sanitize@url \@href}%
\providecommand \@href[1]{\@@startlink{#1}\@@href}%
\providecommand \@@href[1]{\endgroup#1\@@endlink}%
\providecommand \@sanitize@url [0]{\catcode `\\12\catcode `\$12\catcode
  `\&12\catcode `\#12\catcode `\^12\catcode `\_12\catcode `\%12\relax}%
\providecommand \@@startlink[1]{}%
\providecommand \@@endlink[0]{}%
\providecommand \url  [0]{\begingroup\@sanitize@url \@url }%
\providecommand \@url [1]{\endgroup\@href {#1}{\urlprefix }}%
\providecommand \urlprefix  [0]{URL }%
\providecommand \Eprint [0]{\href }%
\providecommand \doibase [0]{http://dx.doi.org/}%
\providecommand \selectlanguage [0]{\@gobble}%
\providecommand \bibinfo  [0]{\@secondoftwo}%
\providecommand \bibfield  [0]{\@secondoftwo}%
\providecommand \translation [1]{[#1]}%
\providecommand \BibitemOpen [0]{}%
\providecommand \bibitemStop [0]{}%
\providecommand \bibitemNoStop [0]{.\EOS\space}%
\providecommand \EOS [0]{\spacefactor3000\relax}%
\providecommand \BibitemShut  [1]{\csname bibitem#1\endcsname}%
\let\auto@bib@innerbib\@empty
%</preamble>
\bibitem [{\citenamefont {Jungman}\ \emph {et~al.}(1996)\citenamefont
  {Jungman}, \citenamefont {Kamionkowski},\ and\ \citenamefont
  {Griest}}]{Jungman:1995df}%
  \BibitemOpen
  \bibfield  {author} {\bibinfo {author} {\bibfnamefont {G.}~\bibnamefont
  {Jungman}}, \bibinfo {author} {\bibfnamefont {M.}~\bibnamefont
  {Kamionkowski}}, \ and\ \bibinfo {author} {\bibfnamefont {K.}~\bibnamefont
  {Griest}},\ }\href {\doibase 10.1016/0370-1573(95)00058-5} {\bibfield
  {journal} {\bibinfo  {journal} {Phys. Rept.}\ }\textbf {\bibinfo {volume}
  {267}},\ \bibinfo {pages} {195} (\bibinfo {year} {1996})},\ \Eprint
  {http://arxiv.org/abs/hep-ph/9506380} {arXiv:hep-ph/9506380 [hep-ph]}
  \BibitemShut {NoStop}%
%%CITATION = HEP-PH/9506380;%%
\bibitem [{\citenamefont {Bertone}\ \emph {et~al.}(2005)\citenamefont
  {Bertone}, \citenamefont {Hooper},\ and\ \citenamefont
  {Silk}}]{Bertone:2004pz}%
  \BibitemOpen
  \bibfield  {author} {\bibinfo {author} {\bibfnamefont {G.}~\bibnamefont
  {Bertone}}, \bibinfo {author} {\bibfnamefont {D.}~\bibnamefont {Hooper}}, \
  and\ \bibinfo {author} {\bibfnamefont {J.}~\bibnamefont {Silk}},\ }\href
  {\doibase 10.1016/j.physrep.2004.08.031} {\bibfield  {journal} {\bibinfo
  {journal} {Phys. Rept.}\ }\textbf {\bibinfo {volume} {405}},\ \bibinfo
  {pages} {279} (\bibinfo {year} {2005})},\ \Eprint
  {http://arxiv.org/abs/hep-ph/0404175} {arXiv:hep-ph/0404175 [hep-ph]}
  \BibitemShut {NoStop}%
%%CITATION = HEP-PH/0404175;%%
\bibitem [{\citenamefont {Goodman}\ and\ \citenamefont
  {Witten}(1985)}]{Goodman:1984dc}%
  \BibitemOpen
  \bibfield  {author} {\bibinfo {author} {\bibfnamefont {M.~W.}\ \bibnamefont
  {Goodman}}\ and\ \bibinfo {author} {\bibfnamefont {E.}~\bibnamefont
  {Witten}},\ }\href {\doibase 10.1103/PhysRevD.31.3059} {\bibfield  {journal}
  {\bibinfo  {journal} {Phys. Rev.}\ }\textbf {\bibinfo {volume} {D31}},\
  \bibinfo {pages} {3059} (\bibinfo {year} {1985})},\ \bibinfo {note}
  {[,325(1984)]}\BibitemShut {NoStop}%
%%CITATION = PHRVA,D31,3059;%%
\bibitem [{\citenamefont {Drukier}\ \emph {et~al.}(1986)\citenamefont
  {Drukier}, \citenamefont {Freese},\ and\ \citenamefont
  {Spergel}}]{Drukier:1986tm}%
  \BibitemOpen
  \bibfield  {author} {\bibinfo {author} {\bibfnamefont {A.~K.}\ \bibnamefont
  {Drukier}}, \bibinfo {author} {\bibfnamefont {K.}~\bibnamefont {Freese}}, \
  and\ \bibinfo {author} {\bibfnamefont {D.~N.}\ \bibnamefont {Spergel}},\
  }\href {\doibase 10.1103/PhysRevD.33.3495} {\bibfield  {journal} {\bibinfo
  {journal} {Phys. Rev.}\ }\textbf {\bibinfo {volume} {D33}},\ \bibinfo {pages}
  {3495} (\bibinfo {year} {1986})}\BibitemShut {NoStop}%
%%CITATION = PHRVA,D33,3495;%%
\bibitem [{\citenamefont {Akerib}\ \emph {et~al.}(2017)\citenamefont {Akerib}
  \emph {et~al.}}]{Akerib:2016vxi}%
  \BibitemOpen
  \bibfield  {author} {\bibinfo {author} {\bibfnamefont {D.~S.}\ \bibnamefont
  {Akerib}} \emph {et~al.} (\bibinfo {collaboration} {LUX}),\ }\href {\doibase
  10.1103/PhysRevLett.118.021303} {\bibfield  {journal} {\bibinfo  {journal}
  {Phys. Rev. Lett.}\ }\textbf {\bibinfo {volume} {118}},\ \bibinfo {pages}
  {021303} (\bibinfo {year} {2017})},\ \Eprint
  {http://arxiv.org/abs/1608.07648} {arXiv:1608.07648 [astro-ph.CO]}
  \BibitemShut {NoStop}%
%%CITATION = ARXIV:1608.07648;%%
\bibitem [{\citenamefont {Aprile}\ \emph {et~al.}(2017)\citenamefont {Aprile}
  \emph {et~al.}}]{Aprile:2017iyp}%
  \BibitemOpen
  \bibfield  {author} {\bibinfo {author} {\bibfnamefont {E.}~\bibnamefont
  {Aprile}} \emph {et~al.} (\bibinfo {collaboration} {XENON}),\ }\href
  {\doibase 10.1103/PhysRevLett.119.181301} {\bibfield  {journal} {\bibinfo
  {journal} {Phys. Rev. Lett.}\ }\textbf {\bibinfo {volume} {119}},\ \bibinfo
  {pages} {181301} (\bibinfo {year} {2017})},\ \Eprint
  {http://arxiv.org/abs/1705.06655} {arXiv:1705.06655 [astro-ph.CO]}
  \BibitemShut {NoStop}%
%%CITATION = ARXIV:1705.06655;%%
\bibitem [{\citenamefont {Cui}\ \emph {et~al.}(2017)\citenamefont {Cui} \emph
  {et~al.}}]{Cui:2017nnn}%
  \BibitemOpen
  \bibfield  {author} {\bibinfo {author} {\bibfnamefont {X.}~\bibnamefont
  {Cui}} \emph {et~al.} (\bibinfo {collaboration} {PandaX-II}),\ }\href
  {\doibase 10.1103/PhysRevLett.119.181302} {\bibfield  {journal} {\bibinfo
  {journal} {Phys. Rev. Lett.}\ }\textbf {\bibinfo {volume} {119}},\ \bibinfo
  {pages} {181302} (\bibinfo {year} {2017})},\ \Eprint
  {http://arxiv.org/abs/1708.06917} {arXiv:1708.06917 [astro-ph.CO]}
  \BibitemShut {NoStop}%
%%CITATION = ARXIV:1708.06917;%%
\bibitem [{\citenamefont {Peter}\ \emph {et~al.}(2014)\citenamefont {Peter},
  \citenamefont {Gluscevic}, \citenamefont {Green}, \citenamefont {Kavanagh},\
  and\ \citenamefont {Lee}}]{Peter:2013aha}%
  \BibitemOpen
  \bibfield  {author} {\bibinfo {author} {\bibfnamefont {A.~H.~G.}\
  \bibnamefont {Peter}}, \bibinfo {author} {\bibfnamefont {V.}~\bibnamefont
  {Gluscevic}}, \bibinfo {author} {\bibfnamefont {A.~M.}\ \bibnamefont
  {Green}}, \bibinfo {author} {\bibfnamefont {B.~J.}\ \bibnamefont {Kavanagh}},
  \ and\ \bibinfo {author} {\bibfnamefont {S.~K.}\ \bibnamefont {Lee}},\ }\href
  {\doibase 10.1016/j.dark.2014.10.006} {\bibfield  {journal} {\bibinfo
  {journal} {Phys. Dark Univ.}\ }\textbf {\bibinfo {volume} {5-6}},\ \bibinfo
  {pages} {45} (\bibinfo {year} {2014})},\ \Eprint
  {http://arxiv.org/abs/1310.7039} {arXiv:1310.7039 [astro-ph.CO]} \BibitemShut
  {NoStop}%
%%CITATION = ARXIV:1310.7039;%%
\bibitem [{\citenamefont {Kavanagh}\ \emph {et~al.}(2015)\citenamefont
  {Kavanagh}, \citenamefont {Fornasa},\ and\ \citenamefont
  {Green}}]{Kavanagh:2014rya}%
  \BibitemOpen
  \bibfield  {author} {\bibinfo {author} {\bibfnamefont {B.~J.}\ \bibnamefont
  {Kavanagh}}, \bibinfo {author} {\bibfnamefont {M.}~\bibnamefont {Fornasa}}, \
  and\ \bibinfo {author} {\bibfnamefont {A.~M.}\ \bibnamefont {Green}},\ }\href
  {\doibase 10.1103/PhysRevD.91.103533} {\bibfield  {journal} {\bibinfo
  {journal} {Phys. Rev.}\ }\textbf {\bibinfo {volume} {D91}},\ \bibinfo {pages}
  {103533} (\bibinfo {year} {2015})},\ \Eprint {http://arxiv.org/abs/1410.8051}
  {arXiv:1410.8051 [astro-ph.CO]} \BibitemShut {NoStop}%
%%CITATION = ARXIV:1410.8051;%%
\bibitem [{\citenamefont {Cahill-Rowley}\ \emph {et~al.}(2015)\citenamefont
  {Cahill-Rowley}, \citenamefont {Cotta}, \citenamefont {Drlica-Wagner},
  \citenamefont {Funk}, \citenamefont {Hewett}, \citenamefont {Ismail},
  \citenamefont {Rizzo},\ and\ \citenamefont {Wood}}]{Cahill-Rowley:2014boa}%
  \BibitemOpen
  \bibfield  {author} {\bibinfo {author} {\bibfnamefont {M.}~\bibnamefont
  {Cahill-Rowley}}, \bibinfo {author} {\bibfnamefont {R.}~\bibnamefont
  {Cotta}}, \bibinfo {author} {\bibfnamefont {A.}~\bibnamefont
  {Drlica-Wagner}}, \bibinfo {author} {\bibfnamefont {S.}~\bibnamefont {Funk}},
  \bibinfo {author} {\bibfnamefont {J.}~\bibnamefont {Hewett}}, \bibinfo
  {author} {\bibfnamefont {A.}~\bibnamefont {Ismail}}, \bibinfo {author}
  {\bibfnamefont {T.}~\bibnamefont {Rizzo}}, \ and\ \bibinfo {author}
  {\bibfnamefont {M.}~\bibnamefont {Wood}},\ }\href {\doibase
  10.1103/PhysRevD.91.055011} {\bibfield  {journal} {\bibinfo  {journal} {Phys.
  Rev.}\ }\textbf {\bibinfo {volume} {D91}},\ \bibinfo {pages} {055011}
  (\bibinfo {year} {2015})},\ \Eprint {http://arxiv.org/abs/1405.6716}
  {arXiv:1405.6716 [hep-ph]} \BibitemShut {NoStop}%
%%CITATION = ARXIV:1405.6716;%%
\bibitem [{\citenamefont {Alves}\ \emph {et~al.}(2015)\citenamefont {Alves},
  \citenamefont {Berlin}, \citenamefont {Profumo},\ and\ \citenamefont
  {Queiroz}}]{Alves:2015pea}%
  \BibitemOpen
  \bibfield  {author} {\bibinfo {author} {\bibfnamefont {A.}~\bibnamefont
  {Alves}}, \bibinfo {author} {\bibfnamefont {A.}~\bibnamefont {Berlin}},
  \bibinfo {author} {\bibfnamefont {S.}~\bibnamefont {Profumo}}, \ and\
  \bibinfo {author} {\bibfnamefont {F.~S.}\ \bibnamefont {Queiroz}},\ }\href
  {\doibase 10.1103/PhysRevD.92.083004} {\bibfield  {journal} {\bibinfo
  {journal} {Phys. Rev.}\ }\textbf {\bibinfo {volume} {D92}},\ \bibinfo {pages}
  {083004} (\bibinfo {year} {2015})},\ \Eprint
  {http://arxiv.org/abs/1501.03490} {arXiv:1501.03490 [hep-ph]} \BibitemShut
  {NoStop}%
%%CITATION = ARXIV:1501.03490;%%
\bibitem [{\citenamefont {Bertone}\ \emph {et~al.}(2018)\citenamefont
  {Bertone}, \citenamefont {Bozorgnia}, \citenamefont {Kim}, \citenamefont
  {Liem}, \citenamefont {McCabe}, \citenamefont {Otten},\ and\ \citenamefont
  {Ruiz~de Austri}}]{Bertone:2017adx}%
  \BibitemOpen
  \bibfield  {author} {\bibinfo {author} {\bibfnamefont {G.}~\bibnamefont
  {Bertone}}, \bibinfo {author} {\bibfnamefont {N.}~\bibnamefont {Bozorgnia}},
  \bibinfo {author} {\bibfnamefont {J.~S.}\ \bibnamefont {Kim}}, \bibinfo
  {author} {\bibfnamefont {S.}~\bibnamefont {Liem}}, \bibinfo {author}
  {\bibfnamefont {C.}~\bibnamefont {McCabe}}, \bibinfo {author} {\bibfnamefont
  {S.}~\bibnamefont {Otten}}, \ and\ \bibinfo {author} {\bibfnamefont
  {R.}~\bibnamefont {Ruiz~de Austri}},\ }\href {\doibase
  10.1088/1475-7516/2018/03/026} {\bibfield  {journal} {\bibinfo  {journal}
  {JCAP}\ }\textbf {\bibinfo {volume} {1803}},\ \bibinfo {pages} {026}
  (\bibinfo {year} {2018})},\ \Eprint {http://arxiv.org/abs/1712.04793}
  {arXiv:1712.04793 [hep-ph]} \BibitemShut {NoStop}%
%%CITATION = ARXIV:1712.04793;%%
\bibitem [{\citenamefont {Catena}(2014{\natexlab{a}})}]{Catena:2014epa}%
  \BibitemOpen
  \bibfield  {author} {\bibinfo {author} {\bibfnamefont {R.}~\bibnamefont
  {Catena}},\ }\href {\doibase 10.1088/1475-7516/2014/07/055} {\bibfield
  {journal} {\bibinfo  {journal} {JCAP}\ }\textbf {\bibinfo {volume} {1407}},\
  \bibinfo {pages} {055} (\bibinfo {year} {2014}{\natexlab{a}})},\ \Eprint
  {http://arxiv.org/abs/1406.0524} {arXiv:1406.0524 [hep-ph]} \BibitemShut
  {NoStop}%
%%CITATION = ARXIV:1406.0524;%%
\bibitem [{\citenamefont {Gresham}\ and\ \citenamefont
  {Zurek}(2014)}]{Gresham:2014vja}%
  \BibitemOpen
  \bibfield  {author} {\bibinfo {author} {\bibfnamefont {M.~I.}\ \bibnamefont
  {Gresham}}\ and\ \bibinfo {author} {\bibfnamefont {K.~M.}\ \bibnamefont
  {Zurek}},\ }\href {\doibase 10.1103/PhysRevD.89.123521} {\bibfield  {journal}
  {\bibinfo  {journal} {Phys. Rev.}\ }\textbf {\bibinfo {volume} {D89}},\
  \bibinfo {pages} {123521} (\bibinfo {year} {2014})},\ \Eprint
  {http://arxiv.org/abs/1401.3739} {arXiv:1401.3739 [hep-ph]} \BibitemShut
  {NoStop}%
%%CITATION = ARXIV:1401.3739;%%
\bibitem [{\citenamefont {Gluscevic}\ and\ \citenamefont
  {Peter}(2014)}]{Gluscevic:2014vga}%
  \BibitemOpen
  \bibfield  {author} {\bibinfo {author} {\bibfnamefont {V.}~\bibnamefont
  {Gluscevic}}\ and\ \bibinfo {author} {\bibfnamefont {A.~H.~G.}\ \bibnamefont
  {Peter}},\ }\href {\doibase 10.1088/1475-7516/2014/09/040} {\bibfield
  {journal} {\bibinfo  {journal} {JCAP}\ }\textbf {\bibinfo {volume} {1409}},\
  \bibinfo {pages} {040} (\bibinfo {year} {2014})},\ \Eprint
  {http://arxiv.org/abs/1406.7008} {arXiv:1406.7008 [astro-ph.CO]} \BibitemShut
  {NoStop}%
%%CITATION = ARXIV:1406.7008;%%
\bibitem [{\citenamefont {Gluscevic}\ \emph {et~al.}(2015)\citenamefont
  {Gluscevic}, \citenamefont {Gresham}, \citenamefont {McDermott},
  \citenamefont {Peter},\ and\ \citenamefont {Zurek}}]{Gluscevic:2015sqa}%
  \BibitemOpen
  \bibfield  {author} {\bibinfo {author} {\bibfnamefont {V.}~\bibnamefont
  {Gluscevic}}, \bibinfo {author} {\bibfnamefont {M.~I.}\ \bibnamefont
  {Gresham}}, \bibinfo {author} {\bibfnamefont {S.~D.}\ \bibnamefont
  {McDermott}}, \bibinfo {author} {\bibfnamefont {A.~H.~G.}\ \bibnamefont
  {Peter}}, \ and\ \bibinfo {author} {\bibfnamefont {K.~M.}\ \bibnamefont
  {Zurek}},\ }\href {\doibase 10.1088/1475-7516/2015/12/057} {\bibfield
  {journal} {\bibinfo  {journal} {JCAP}\ }\textbf {\bibinfo {volume} {1512}},\
  \bibinfo {pages} {057} (\bibinfo {year} {2015})},\ \Eprint
  {http://arxiv.org/abs/1506.04454} {arXiv:1506.04454 [hep-ph]} \BibitemShut
  {NoStop}%
%%CITATION = ARXIV:1506.04454;%%
\bibitem [{\citenamefont {Kahlhoefer}\ and\ \citenamefont
  {Wild}(2016)}]{Kahlhoefer:2016eds}%
  \BibitemOpen
  \bibfield  {author} {\bibinfo {author} {\bibfnamefont {F.}~\bibnamefont
  {Kahlhoefer}}\ and\ \bibinfo {author} {\bibfnamefont {S.}~\bibnamefont
  {Wild}},\ }\href {\doibase 10.1088/1475-7516/2016/10/032} {\bibfield
  {journal} {\bibinfo  {journal} {JCAP}\ }\textbf {\bibinfo {volume} {1610}},\
  \bibinfo {pages} {032} (\bibinfo {year} {2016})},\ \Eprint
  {http://arxiv.org/abs/1607.04418} {arXiv:1607.04418 [hep-ph]} \BibitemShut
  {NoStop}%
%%CITATION = ARXIV:1607.04418;%%
\bibitem [{\citenamefont {Catena}\ \emph {et~al.}(2017)\citenamefont {Catena},
  \citenamefont {Conrad}, \citenamefont {Döring}, \citenamefont {Ferella},\
  and\ \citenamefont {Krauss}}]{Catena:2017wzu}%
  \BibitemOpen
  \bibfield  {author} {\bibinfo {author} {\bibfnamefont {R.}~\bibnamefont
  {Catena}}, \bibinfo {author} {\bibfnamefont {J.}~\bibnamefont {Conrad}},
  \bibinfo {author} {\bibfnamefont {C.}~\bibnamefont {Döring}}, \bibinfo
  {author} {\bibfnamefont {A.~D.}\ \bibnamefont {Ferella}}, \ and\ \bibinfo
  {author} {\bibfnamefont {M.~B.}\ \bibnamefont {Krauss}},\ }\href@noop {} {\
  (\bibinfo {year} {2017})},\ \Eprint {http://arxiv.org/abs/1706.09471}
  {arXiv:1706.09471 [hep-ph]} \BibitemShut {NoStop}%
%%CITATION = ARXIV:1706.09471;%%
\bibitem [{\citenamefont {Baum}\ \emph {et~al.}(2017)\citenamefont {Baum},
  \citenamefont {Catena}, \citenamefont {Conrad}, \citenamefont {Freese},\ and\
  \citenamefont {Krauss}}]{Baum:2017kfa}%
  \BibitemOpen
  \bibfield  {author} {\bibinfo {author} {\bibfnamefont {S.}~\bibnamefont
  {Baum}}, \bibinfo {author} {\bibfnamefont {R.}~\bibnamefont {Catena}},
  \bibinfo {author} {\bibfnamefont {J.}~\bibnamefont {Conrad}}, \bibinfo
  {author} {\bibfnamefont {K.}~\bibnamefont {Freese}}, \ and\ \bibinfo {author}
  {\bibfnamefont {M.~B.}\ \bibnamefont {Krauss}},\ }\href@noop {} {\  (\bibinfo
  {year} {2017})},\ \Eprint {http://arxiv.org/abs/1709.06051} {arXiv:1709.06051
  [hep-ph]} \BibitemShut {NoStop}%
%%CITATION = ARXIV:1709.06051;%%
\bibitem [{\citenamefont {Fan}\ \emph {et~al.}(2010)\citenamefont {Fan},
  \citenamefont {Reece},\ and\ \citenamefont {Wang}}]{Fan:2010gt}%
  \BibitemOpen
  \bibfield  {author} {\bibinfo {author} {\bibfnamefont {J.}~\bibnamefont
  {Fan}}, \bibinfo {author} {\bibfnamefont {M.}~\bibnamefont {Reece}}, \ and\
  \bibinfo {author} {\bibfnamefont {L.-T.}\ \bibnamefont {Wang}},\ }\href
  {\doibase 10.1088/1475-7516/2010/11/042} {\bibfield  {journal} {\bibinfo
  {journal} {JCAP}\ }\textbf {\bibinfo {volume} {1011}},\ \bibinfo {pages}
  {042} (\bibinfo {year} {2010})},\ \Eprint {http://arxiv.org/abs/1008.1591}
  {arXiv:1008.1591 [hep-ph]} \BibitemShut {NoStop}%
%%CITATION = ARXIV:1008.1591;%%
\bibitem [{\citenamefont {Fitzpatrick}\ \emph {et~al.}(2013)\citenamefont
  {Fitzpatrick}, \citenamefont {Haxton}, \citenamefont {Katz}, \citenamefont
  {Lubbers},\ and\ \citenamefont {Xu}}]{Fitzpatrick:2012ix}%
  \BibitemOpen
  \bibfield  {author} {\bibinfo {author} {\bibfnamefont {A.~L.}\ \bibnamefont
  {Fitzpatrick}}, \bibinfo {author} {\bibfnamefont {W.}~\bibnamefont {Haxton}},
  \bibinfo {author} {\bibfnamefont {E.}~\bibnamefont {Katz}}, \bibinfo {author}
  {\bibfnamefont {N.}~\bibnamefont {Lubbers}}, \ and\ \bibinfo {author}
  {\bibfnamefont {Y.}~\bibnamefont {Xu}},\ }\href {\doibase
  10.1088/1475-7516/2013/02/004} {\bibfield  {journal} {\bibinfo  {journal}
  {JCAP}\ }\textbf {\bibinfo {volume} {1302}},\ \bibinfo {pages} {004}
  (\bibinfo {year} {2013})},\ \Eprint {http://arxiv.org/abs/1203.3542}
  {arXiv:1203.3542 [hep-ph]} \BibitemShut {NoStop}%
%%CITATION = ARXIV:1203.3542;%%
\bibitem [{\citenamefont {Fitzpatrick}\ \emph {et~al.}(2012)\citenamefont
  {Fitzpatrick}, \citenamefont {Haxton}, \citenamefont {Katz}, \citenamefont
  {Lubbers},\ and\ \citenamefont {Xu}}]{Fitzpatrick:2012ib}%
  \BibitemOpen
  \bibfield  {author} {\bibinfo {author} {\bibfnamefont {A.~L.}\ \bibnamefont
  {Fitzpatrick}}, \bibinfo {author} {\bibfnamefont {W.}~\bibnamefont {Haxton}},
  \bibinfo {author} {\bibfnamefont {E.}~\bibnamefont {Katz}}, \bibinfo {author}
  {\bibfnamefont {N.}~\bibnamefont {Lubbers}}, \ and\ \bibinfo {author}
  {\bibfnamefont {Y.}~\bibnamefont {Xu}},\ }\href@noop {} {\  (\bibinfo {year}
  {2012})},\ \Eprint {http://arxiv.org/abs/1211.2818} {arXiv:1211.2818
  [hep-ph]} \BibitemShut {NoStop}%
%%CITATION = ARXIV:1211.2818;%%
\bibitem [{\citenamefont {Anand}\ \emph {et~al.}(2014)\citenamefont {Anand},
  \citenamefont {Fitzpatrick},\ and\ \citenamefont {Haxton}}]{Anand:2013yka}%
  \BibitemOpen
  \bibfield  {author} {\bibinfo {author} {\bibfnamefont {N.}~\bibnamefont
  {Anand}}, \bibinfo {author} {\bibfnamefont {A.~L.}\ \bibnamefont
  {Fitzpatrick}}, \ and\ \bibinfo {author} {\bibfnamefont {W.~C.}\ \bibnamefont
  {Haxton}},\ }\href {\doibase 10.1103/PhysRevC.89.065501} {\bibfield
  {journal} {\bibinfo  {journal} {Phys. Rev.}\ }\textbf {\bibinfo {volume}
  {C89}},\ \bibinfo {pages} {065501} (\bibinfo {year} {2014})},\ \Eprint
  {http://arxiv.org/abs/1308.6288} {arXiv:1308.6288 [hep-ph]} \BibitemShut
  {NoStop}%
%%CITATION = ARXIV:1308.6288;%%
\bibitem [{\citenamefont {Dent}\ \emph {et~al.}(2015)\citenamefont {Dent},
  \citenamefont {Krauss}, \citenamefont {Newstead},\ and\ \citenamefont
  {Sabharwal}}]{Dent:2015zpa}%
  \BibitemOpen
  \bibfield  {author} {\bibinfo {author} {\bibfnamefont {J.~B.}\ \bibnamefont
  {Dent}}, \bibinfo {author} {\bibfnamefont {L.~M.}\ \bibnamefont {Krauss}},
  \bibinfo {author} {\bibfnamefont {J.~L.}\ \bibnamefont {Newstead}}, \ and\
  \bibinfo {author} {\bibfnamefont {S.}~\bibnamefont {Sabharwal}},\ }\href
  {\doibase 10.1103/PhysRevD.92.063515} {\bibfield  {journal} {\bibinfo
  {journal} {Phys. Rev.}\ }\textbf {\bibinfo {volume} {D92}},\ \bibinfo {pages}
  {063515} (\bibinfo {year} {2015})},\ \Eprint
  {http://arxiv.org/abs/1505.03117} {arXiv:1505.03117 [hep-ph]} \BibitemShut
  {NoStop}%
%%CITATION = ARXIV:1505.03117;%%
\bibitem [{\citenamefont {Catena}(2014{\natexlab{b}})}]{Catena:2014hla}%
  \BibitemOpen
  \bibfield  {author} {\bibinfo {author} {\bibfnamefont {R.}~\bibnamefont
  {Catena}},\ }\href {\doibase 10.1088/1475-7516/2014/09/049} {\bibfield
  {journal} {\bibinfo  {journal} {JCAP}\ }\textbf {\bibinfo {volume} {1409}},\
  \bibinfo {pages} {049} (\bibinfo {year} {2014}{\natexlab{b}})},\ \Eprint
  {http://arxiv.org/abs/1407.0127} {arXiv:1407.0127 [hep-ph]} \BibitemShut
  {NoStop}%
%%CITATION = ARXIV:1407.0127;%%
\bibitem [{\citenamefont {Catena}\ and\ \citenamefont
  {Gondolo}(2014)}]{Catena:2014uqa}%
  \BibitemOpen
  \bibfield  {author} {\bibinfo {author} {\bibfnamefont {R.}~\bibnamefont
  {Catena}}\ and\ \bibinfo {author} {\bibfnamefont {P.}~\bibnamefont
  {Gondolo}},\ }\href {\doibase 10.1088/1475-7516/2014/09/045} {\bibfield
  {journal} {\bibinfo  {journal} {JCAP}\ }\textbf {\bibinfo {volume} {1409}},\
  \bibinfo {pages} {045} (\bibinfo {year} {2014})},\ \Eprint
  {http://arxiv.org/abs/1405.2637} {arXiv:1405.2637 [hep-ph]} \BibitemShut
  {NoStop}%
%%CITATION = ARXIV:1405.2637;%%
\bibitem [{\citenamefont {Catena}\ and\ \citenamefont
  {Gondolo}(2015)}]{Catena:2015uua}%
  \BibitemOpen
  \bibfield  {author} {\bibinfo {author} {\bibfnamefont {R.}~\bibnamefont
  {Catena}}\ and\ \bibinfo {author} {\bibfnamefont {P.}~\bibnamefont
  {Gondolo}},\ }\href {\doibase 10.1088/1475-7516/2015/08/022} {\bibfield
  {journal} {\bibinfo  {journal} {JCAP}\ }\textbf {\bibinfo {volume} {1508}},\
  \bibinfo {pages} {022} (\bibinfo {year} {2015})},\ \Eprint
  {http://arxiv.org/abs/1504.06554} {arXiv:1504.06554 [hep-ph]} \BibitemShut
  {NoStop}%
%%CITATION = ARXIV:1504.06554;%%
\bibitem [{\citenamefont {Cowan}\ \emph {et~al.}(2011)\citenamefont {Cowan},
  \citenamefont {Cranmer}, \citenamefont {Gross},\ and\ \citenamefont
  {Vitells}}]{Cowan:2010js}%
  \BibitemOpen
  \bibfield  {author} {\bibinfo {author} {\bibfnamefont {G.}~\bibnamefont
  {Cowan}}, \bibinfo {author} {\bibfnamefont {K.}~\bibnamefont {Cranmer}},
  \bibinfo {author} {\bibfnamefont {E.}~\bibnamefont {Gross}}, \ and\ \bibinfo
  {author} {\bibfnamefont {O.}~\bibnamefont {Vitells}},\ }\href {\doibase
  10.1140/epjc/s10052-011-1554-0, 10.1140/epjc/s10052-013-2501-z} {\bibfield
  {journal} {\bibinfo  {journal} {Eur. Phys. J.}\ }\textbf {\bibinfo {volume}
  {C71}},\ \bibinfo {pages} {1554} (\bibinfo {year} {2011})},\ \bibinfo {note}
  {[Erratum: Eur. Phys. J.C73,2501(2013)]},\ \Eprint
  {http://arxiv.org/abs/1007.1727} {arXiv:1007.1727 [physics.data-an]}
  \BibitemShut {NoStop}%
%%CITATION = ARXIV:1007.1727;%%
\bibitem [{\citenamefont {Wilks}(1938)}]{Wilks:1938dza}%
  \BibitemOpen
  \bibfield  {author} {\bibinfo {author} {\bibfnamefont {S.~S.}\ \bibnamefont
  {Wilks}},\ }\href {\doibase 10.1214/aoms/1177732360} {\bibfield  {journal}
  {\bibinfo  {journal} {Annals Math. Statist.}\ }\textbf {\bibinfo {volume}
  {9}},\ \bibinfo {pages} {60} (\bibinfo {year} {1938})}\BibitemShut {NoStop}%
%%CITATION = AASTA,9,60;%%
\bibitem [{\citenamefont {Edwards}\ and\ \citenamefont
  {Weniger}(2017)}]{Edwards:2017kqw}%
  \BibitemOpen
  \bibfield  {author} {\bibinfo {author} {\bibfnamefont {T.~D.~P.}\
  \bibnamefont {Edwards}}\ and\ \bibinfo {author} {\bibfnamefont
  {C.}~\bibnamefont {Weniger}},\ }\href@noop {} {\  (\bibinfo {year} {2017})},\
  \Eprint {http://arxiv.org/abs/1712.05401} {arXiv:1712.05401 [hep-ph]}
  \BibitemShut {NoStop}%
%%CITATION = ARXIV:1712.05401;%%
\bibitem [{\citenamefont {Venn}(1880)}]{Venn1880}%
  \BibitemOpen
  \bibfield  {author} {\bibinfo {author} {\bibfnamefont {J.}~\bibnamefont
  {Venn}},\ }\href {\doibase 10.1080/14786448008626877} {\bibfield  {journal}
  {\bibinfo  {journal} {The London, Edinburgh, and Dublin Philosophical
  Magazine and Journal of Science}\ }\textbf {\bibinfo {volume} {10}},\
  \bibinfo {pages} {1} (\bibinfo {year} {1880})}\BibitemShut {NoStop}%
\bibitem [{\citenamefont {Cerdeno}\ and\ \citenamefont
  {Green}(2010)}]{Cerdeno:2010jj}%
  \BibitemOpen
  \bibfield  {author} {\bibinfo {author} {\bibfnamefont {D.~G.}\ \bibnamefont
  {Cerdeno}}\ and\ \bibinfo {author} {\bibfnamefont {A.~M.}\ \bibnamefont
  {Green}},\ }\enquote {\bibinfo {title} {{Direct detection of WIMPs}},}\ in\
  \href@noop {} {\emph {\bibinfo {booktitle} {Particle Dark Matter:
  Observations, Models and Searches}}},\ \bibinfo {editor} {edited by\ \bibinfo
  {editor} {\bibfnamefont {G.}~\bibnamefont {Bertone}}}\ (\bibinfo {year}
  {2010})\ pp.\ \bibinfo {pages} {347--369},\ \Eprint
  {http://arxiv.org/abs/1002.1912} {arXiv:1002.1912 [astro-ph.CO]} \BibitemShut
  {NoStop}%
%%CITATION = ARXIV:1002.1912;%%
\bibitem [{\citenamefont {Chang}\ \emph {et~al.}(2010)\citenamefont {Chang},
  \citenamefont {Pierce},\ and\ \citenamefont {Weiner}}]{Chang:2009yt}%
  \BibitemOpen
  \bibfield  {author} {\bibinfo {author} {\bibfnamefont {S.}~\bibnamefont
  {Chang}}, \bibinfo {author} {\bibfnamefont {A.}~\bibnamefont {Pierce}}, \
  and\ \bibinfo {author} {\bibfnamefont {N.}~\bibnamefont {Weiner}},\ }\href
  {\doibase 10.1088/1475-7516/2010/01/006} {\bibfield  {journal} {\bibinfo
  {journal} {JCAP}\ }\textbf {\bibinfo {volume} {1001}},\ \bibinfo {pages}
  {006} (\bibinfo {year} {2010})},\ \Eprint {http://arxiv.org/abs/0908.3192}
  {arXiv:0908.3192 [hep-ph]} \BibitemShut {NoStop}%
%%CITATION = ARXIV:0908.3192;%%
\bibitem [{\citenamefont {Catena}(2015)}]{Catena:2015vpa}%
  \BibitemOpen
  \bibfield  {author} {\bibinfo {author} {\bibfnamefont {R.}~\bibnamefont
  {Catena}},\ }\href {\doibase 10.1088/1475-7516/2015/07/026} {\bibfield
  {journal} {\bibinfo  {journal} {JCAP}\ }\textbf {\bibinfo {volume} {1507}},\
  \bibinfo {pages} {026} (\bibinfo {year} {2015})},\ \Eprint
  {http://arxiv.org/abs/1505.06441} {arXiv:1505.06441 [hep-ph]} \BibitemShut
  {NoStop}%
%%CITATION = ARXIV:1505.06441;%%
\bibitem [{\citenamefont {Kavanagh}(2015)}]{Kavanagh:2015jma}%
  \BibitemOpen
  \bibfield  {author} {\bibinfo {author} {\bibfnamefont {B.~J.}\ \bibnamefont
  {Kavanagh}},\ }\href {\doibase 10.1103/PhysRevD.92.023513} {\bibfield
  {journal} {\bibinfo  {journal} {Phys. Rev.}\ }\textbf {\bibinfo {volume}
  {D92}},\ \bibinfo {pages} {023513} (\bibinfo {year} {2015})},\ \Eprint
  {http://arxiv.org/abs/1505.07406} {arXiv:1505.07406 [hep-ph]} \BibitemShut
  {NoStop}%
%%CITATION = ARXIV:1505.07406;%%
\bibitem [{\citenamefont {Del~Nobile}\ \emph {et~al.}(2015)\citenamefont
  {Del~Nobile}, \citenamefont {Gelmini},\ and\ \citenamefont
  {Witte}}]{DelNobile:2015tza}%
  \BibitemOpen
  \bibfield  {author} {\bibinfo {author} {\bibfnamefont {E.}~\bibnamefont
  {Del~Nobile}}, \bibinfo {author} {\bibfnamefont {G.~B.}\ \bibnamefont
  {Gelmini}}, \ and\ \bibinfo {author} {\bibfnamefont {S.~J.}\ \bibnamefont
  {Witte}},\ }\href {\doibase 10.1103/PhysRevD.91.121302} {\bibfield  {journal}
  {\bibinfo  {journal} {Phys. Rev.}\ }\textbf {\bibinfo {volume} {D91}},\
  \bibinfo {pages} {121302} (\bibinfo {year} {2015})},\ \Eprint
  {http://arxiv.org/abs/1504.06772} {arXiv:1504.06772 [hep-ph]} \BibitemShut
  {NoStop}%
%%CITATION = ARXIV:1504.06772;%%
\bibitem [{\citenamefont {Del~Nobile}\ \emph {et~al.}(2016)\citenamefont
  {Del~Nobile}, \citenamefont {Gelmini},\ and\ \citenamefont
  {Witte}}]{DelNobile:2015rmp}%
  \BibitemOpen
  \bibfield  {author} {\bibinfo {author} {\bibfnamefont {E.}~\bibnamefont
  {Del~Nobile}}, \bibinfo {author} {\bibfnamefont {G.~B.}\ \bibnamefont
  {Gelmini}}, \ and\ \bibinfo {author} {\bibfnamefont {S.~J.}\ \bibnamefont
  {Witte}},\ }\href {\doibase 10.1088/1475-7516/2016/02/009} {\bibfield
  {journal} {\bibinfo  {journal} {JCAP}\ }\textbf {\bibinfo {volume} {1602}},\
  \bibinfo {pages} {009} (\bibinfo {year} {2016})},\ \Eprint
  {http://arxiv.org/abs/1512.03961} {arXiv:1512.03961 [hep-ph]} \BibitemShut
  {NoStop}%
%%CITATION = ARXIV:1512.03961;%%
\bibitem [{\citenamefont {D'Eramo}\ \emph {et~al.}(2016)\citenamefont
  {D'Eramo}, \citenamefont {Kavanagh},\ and\ \citenamefont
  {Panci}}]{DEramo:2016gos}%
  \BibitemOpen
  \bibfield  {author} {\bibinfo {author} {\bibfnamefont {F.}~\bibnamefont
  {D'Eramo}}, \bibinfo {author} {\bibfnamefont {B.~J.}\ \bibnamefont
  {Kavanagh}}, \ and\ \bibinfo {author} {\bibfnamefont {P.}~\bibnamefont
  {Panci}},\ }\href {\doibase 10.1007/JHEP08(2016)111} {\bibfield  {journal}
  {\bibinfo  {journal} {JHEP}\ }\textbf {\bibinfo {volume} {08}},\ \bibinfo
  {pages} {111} (\bibinfo {year} {2016})},\ \Eprint
  {http://arxiv.org/abs/1605.04917} {arXiv:1605.04917 [hep-ph]} \BibitemShut
  {NoStop}%
%%CITATION = ARXIV:1605.04917;%%
\bibitem [{\citenamefont {Bishara}\ \emph {et~al.}(2017)\citenamefont
  {Bishara}, \citenamefont {Brod}, \citenamefont {Grinstein},\ and\
  \citenamefont {Zupan}}]{Bishara:2017nnn}%
  \BibitemOpen
  \bibfield  {author} {\bibinfo {author} {\bibfnamefont {F.}~\bibnamefont
  {Bishara}}, \bibinfo {author} {\bibfnamefont {J.}~\bibnamefont {Brod}},
  \bibinfo {author} {\bibfnamefont {B.}~\bibnamefont {Grinstein}}, \ and\
  \bibinfo {author} {\bibfnamefont {J.}~\bibnamefont {Zupan}},\ }\href@noop {}
  {\  (\bibinfo {year} {2017})},\ \Eprint {http://arxiv.org/abs/1708.02678}
  {arXiv:1708.02678 [hep-ph]} \BibitemShut {NoStop}%
%%CITATION = ARXIV:1708.02678;%%
\bibitem [{\citenamefont {Cerdeño}\ \emph {et~al.}(2013)\citenamefont
  {Cerdeño} \emph {et~al.}}]{Cerdeno:2013gqa}%
  \BibitemOpen
  \bibfield  {author} {\bibinfo {author} {\bibfnamefont {D.~G.}\ \bibnamefont
  {Cerdeño}} \emph {et~al.},\ }\href {\doibase 10.1088/1475-7516/2013/07/028,
  10.1088/1475-7516/2013/09/E01} {\bibfield  {journal} {\bibinfo  {journal}
  {JCAP}\ }\textbf {\bibinfo {volume} {1307}},\ \bibinfo {pages} {028}
  (\bibinfo {year} {2013})},\ \bibinfo {note} {[Erratum: JCAP1309,E01(2013)]},\
  \Eprint {http://arxiv.org/abs/1304.1758} {arXiv:1304.1758 [hep-ph]}
  \BibitemShut {NoStop}%
%%CITATION = ARXIV:1304.1758;%%
\bibitem [{\citenamefont {Cirelli}\ \emph {et~al.}(2013)\citenamefont
  {Cirelli}, \citenamefont {Del~Nobile},\ and\ \citenamefont
  {Panci}}]{DelNobile:2013sia}%
  \BibitemOpen
  \bibfield  {author} {\bibinfo {author} {\bibfnamefont {M.}~\bibnamefont
  {Cirelli}}, \bibinfo {author} {\bibfnamefont {E.}~\bibnamefont {Del~Nobile}},
  \ and\ \bibinfo {author} {\bibfnamefont {P.}~\bibnamefont {Panci}},\ }\href
  {\doibase 10.1088/1475-7516/2013/10/019} {\bibfield  {journal} {\bibinfo
  {journal} {JCAP}\ }\textbf {\bibinfo {volume} {1310}},\ \bibinfo {pages}
  {019} (\bibinfo {year} {2013})},\ \Eprint {http://arxiv.org/abs/1307.5955}
  {arXiv:1307.5955 [hep-ph]} \BibitemShut {NoStop}%
%%CITATION = ARXIV:1307.5955;%%
\bibitem [{\citenamefont {Kahlhoefer}\ \emph {et~al.}(2017)\citenamefont
  {Kahlhoefer}, \citenamefont {Kulkarni},\ and\ \citenamefont
  {Wild}}]{Kahlhoefer:2017ddj}%
  \BibitemOpen
  \bibfield  {author} {\bibinfo {author} {\bibfnamefont {F.}~\bibnamefont
  {Kahlhoefer}}, \bibinfo {author} {\bibfnamefont {S.}~\bibnamefont
  {Kulkarni}}, \ and\ \bibinfo {author} {\bibfnamefont {S.}~\bibnamefont
  {Wild}},\ }\href {\doibase 10.1088/1475-7516/2017/11/016} {\bibfield
  {journal} {\bibinfo  {journal} {JCAP}\ }\textbf {\bibinfo {volume} {1711}},\
  \bibinfo {pages} {016} (\bibinfo {year} {2017})},\ \Eprint
  {http://arxiv.org/abs/1707.08571} {arXiv:1707.08571 [hep-ph]} \BibitemShut
  {NoStop}%
%%CITATION = ARXIV:1707.08571;%%
\bibitem [{\citenamefont {Ho}\ and\ \citenamefont
  {Scherrer}(2013)}]{Ho:2012bg}%
  \BibitemOpen
  \bibfield  {author} {\bibinfo {author} {\bibfnamefont {C.~M.}\ \bibnamefont
  {Ho}}\ and\ \bibinfo {author} {\bibfnamefont {R.~J.}\ \bibnamefont
  {Scherrer}},\ }\href {\doibase 10.1016/j.physletb.2013.04.039} {\bibfield
  {journal} {\bibinfo  {journal} {Phys. Lett.}\ }\textbf {\bibinfo {volume}
  {B722}},\ \bibinfo {pages} {341} (\bibinfo {year} {2013})},\ \Eprint
  {http://arxiv.org/abs/1211.0503} {arXiv:1211.0503 [hep-ph]} \BibitemShut
  {NoStop}%
%%CITATION = ARXIV:1211.0503;%%
\bibitem [{\citenamefont {Del~Nobile}\ \emph {et~al.}(2014)\citenamefont
  {Del~Nobile}, \citenamefont {Gelmini}, \citenamefont {Gondolo},\ and\
  \citenamefont {Huh}}]{DelNobile:2014eta}%
  \BibitemOpen
  \bibfield  {author} {\bibinfo {author} {\bibfnamefont {E.}~\bibnamefont
  {Del~Nobile}}, \bibinfo {author} {\bibfnamefont {G.~B.}\ \bibnamefont
  {Gelmini}}, \bibinfo {author} {\bibfnamefont {P.}~\bibnamefont {Gondolo}}, \
  and\ \bibinfo {author} {\bibfnamefont {J.-H.}\ \bibnamefont {Huh}},\ }\href
  {\doibase 10.1088/1475-7516/2014/06/002} {\bibfield  {journal} {\bibinfo
  {journal} {JCAP}\ }\textbf {\bibinfo {volume} {1406}},\ \bibinfo {pages}
  {002} (\bibinfo {year} {2014})},\ \Eprint {http://arxiv.org/abs/1401.4508}
  {arXiv:1401.4508 [hep-ph]} \BibitemShut {NoStop}%
%%CITATION = ARXIV:1401.4508;%%
\bibitem [{\citenamefont {McDermott}\ \emph {et~al.}(2011)\citenamefont
  {McDermott}, \citenamefont {Yu},\ and\ \citenamefont
  {Zurek}}]{McDermott:2010pa}%
  \BibitemOpen
  \bibfield  {author} {\bibinfo {author} {\bibfnamefont {S.~D.}\ \bibnamefont
  {McDermott}}, \bibinfo {author} {\bibfnamefont {H.-B.}\ \bibnamefont {Yu}}, \
  and\ \bibinfo {author} {\bibfnamefont {K.~M.}\ \bibnamefont {Zurek}},\ }\href
  {\doibase 10.1103/PhysRevD.83.063509} {\bibfield  {journal} {\bibinfo
  {journal} {Phys. Rev.}\ }\textbf {\bibinfo {volume} {D83}},\ \bibinfo {pages}
  {063509} (\bibinfo {year} {2011})},\ \Eprint {http://arxiv.org/abs/1011.2907}
  {arXiv:1011.2907 [hep-ph]} \BibitemShut {NoStop}%
%%CITATION = ARXIV:1011.2907;%%
\bibitem [{\citenamefont {Fornengo}\ \emph {et~al.}(2011)\citenamefont
  {Fornengo}, \citenamefont {Panci},\ and\ \citenamefont
  {Regis}}]{Fornengo:2011sz}%
  \BibitemOpen
  \bibfield  {author} {\bibinfo {author} {\bibfnamefont {N.}~\bibnamefont
  {Fornengo}}, \bibinfo {author} {\bibfnamefont {P.}~\bibnamefont {Panci}}, \
  and\ \bibinfo {author} {\bibfnamefont {M.}~\bibnamefont {Regis}},\ }\href
  {\doibase 10.1103/PhysRevD.84.115002} {\bibfield  {journal} {\bibinfo
  {journal} {Phys. Rev.}\ }\textbf {\bibinfo {volume} {D84}},\ \bibinfo {pages}
  {115002} (\bibinfo {year} {2011})},\ \Eprint {http://arxiv.org/abs/1108.4661}
  {arXiv:1108.4661 [hep-ph]} \BibitemShut {NoStop}%
%%CITATION = ARXIV:1108.4661;%%
\bibitem [{\citenamefont {Panci}(2014)}]{Panci:2014gga}%
  \BibitemOpen
  \bibfield  {author} {\bibinfo {author} {\bibfnamefont {P.}~\bibnamefont
  {Panci}},\ }\href {\doibase 10.1155/2014/681312} {\bibfield  {journal}
  {\bibinfo  {journal} {Adv. High Energy Phys.}\ }\textbf {\bibinfo {volume}
  {2014}},\ \bibinfo {pages} {681312} (\bibinfo {year} {2014})},\ \Eprint
  {http://arxiv.org/abs/1402.1507} {arXiv:1402.1507 [hep-ph]} \BibitemShut
  {NoStop}%
%%CITATION = ARXIV:1402.1507;%%
\bibitem [{\citenamefont {Sigurdson}\ \emph {et~al.}(2004)\citenamefont
  {Sigurdson}, \citenamefont {Doran}, \citenamefont {Kurylov}, \citenamefont
  {Caldwell},\ and\ \citenamefont {Kamionkowski}}]{Sigurdson:2004zp}%
  \BibitemOpen
  \bibfield  {author} {\bibinfo {author} {\bibfnamefont {K.}~\bibnamefont
  {Sigurdson}}, \bibinfo {author} {\bibfnamefont {M.}~\bibnamefont {Doran}},
  \bibinfo {author} {\bibfnamefont {A.}~\bibnamefont {Kurylov}}, \bibinfo
  {author} {\bibfnamefont {R.~R.}\ \bibnamefont {Caldwell}}, \ and\ \bibinfo
  {author} {\bibfnamefont {M.}~\bibnamefont {Kamionkowski}},\ }\href {\doibase
  10.1103/PhysRevD.70.083501, 10.1103/PhysRevD.73.089903} {\bibfield  {journal}
  {\bibinfo  {journal} {Phys. Rev.}\ }\textbf {\bibinfo {volume} {D70}},\
  \bibinfo {pages} {083501} (\bibinfo {year} {2004})},\ \bibinfo {note}
  {[Erratum: Phys. Rev.D73,089903(2006)]},\ \Eprint
  {http://arxiv.org/abs/astro-ph/0406355} {arXiv:astro-ph/0406355 [astro-ph]}
  \BibitemShut {NoStop}%
%%CITATION = ASTRO-PH/0406355;%%
\bibitem [{\citenamefont {Banks}\ \emph {et~al.}(2010)\citenamefont {Banks},
  \citenamefont {Fortin},\ and\ \citenamefont {Thomas}}]{Banks:2010eh}%
  \BibitemOpen
  \bibfield  {author} {\bibinfo {author} {\bibfnamefont {T.}~\bibnamefont
  {Banks}}, \bibinfo {author} {\bibfnamefont {J.-F.}\ \bibnamefont {Fortin}}, \
  and\ \bibinfo {author} {\bibfnamefont {S.}~\bibnamefont {Thomas}},\
  }\href@noop {} {\  (\bibinfo {year} {2010})},\ \Eprint
  {http://arxiv.org/abs/1007.5515} {arXiv:1007.5515 [hep-ph]} \BibitemShut
  {NoStop}%
%%CITATION = ARXIV:1007.5515;%%
\bibitem [{\citenamefont {Bagnasco}\ \emph {et~al.}(1994)\citenamefont
  {Bagnasco}, \citenamefont {Dine},\ and\ \citenamefont
  {Thomas}}]{Bagnasco:1993st}%
  \BibitemOpen
  \bibfield  {author} {\bibinfo {author} {\bibfnamefont {J.}~\bibnamefont
  {Bagnasco}}, \bibinfo {author} {\bibfnamefont {M.}~\bibnamefont {Dine}}, \
  and\ \bibinfo {author} {\bibfnamefont {S.~D.}\ \bibnamefont {Thomas}},\
  }\href {\doibase 10.1016/0370-2693(94)90830-3} {\bibfield  {journal}
  {\bibinfo  {journal} {Phys. Lett.}\ }\textbf {\bibinfo {volume} {B320}},\
  \bibinfo {pages} {99} (\bibinfo {year} {1994})},\ \Eprint
  {http://arxiv.org/abs/hep-ph/9310290} {arXiv:hep-ph/9310290 [hep-ph]}
  \BibitemShut {NoStop}%
%%CITATION = HEP-PH/9310290;%%
\bibitem [{\citenamefont {Banks}\ \emph {et~al.}(2005)\citenamefont {Banks},
  \citenamefont {Mason},\ and\ \citenamefont {O'Neil}}]{Banks:2005hc}%
  \BibitemOpen
  \bibfield  {author} {\bibinfo {author} {\bibfnamefont {T.}~\bibnamefont
  {Banks}}, \bibinfo {author} {\bibfnamefont {J.~D.}\ \bibnamefont {Mason}}, \
  and\ \bibinfo {author} {\bibfnamefont {D.}~\bibnamefont {O'Neil}},\ }\href
  {\doibase 10.1103/PhysRevD.72.043530} {\bibfield  {journal} {\bibinfo
  {journal} {Phys. Rev.}\ }\textbf {\bibinfo {volume} {D72}},\ \bibinfo {pages}
  {043530} (\bibinfo {year} {2005})},\ \Eprint
  {http://arxiv.org/abs/hep-ph/0506015} {arXiv:hep-ph/0506015 [hep-ph]}
  \BibitemShut {NoStop}%
%%CITATION = HEP-PH/0506015;%%
\bibitem [{\citenamefont {Barger}\ \emph {et~al.}(2011)\citenamefont {Barger},
  \citenamefont {Keung},\ and\ \citenamefont {Marfatia}}]{Barger:2010gv}%
  \BibitemOpen
  \bibfield  {author} {\bibinfo {author} {\bibfnamefont {V.}~\bibnamefont
  {Barger}}, \bibinfo {author} {\bibfnamefont {W.-Y.}\ \bibnamefont {Keung}}, \
  and\ \bibinfo {author} {\bibfnamefont {D.}~\bibnamefont {Marfatia}},\ }\href
  {\doibase 10.1016/j.physletb.2010.12.008} {\bibfield  {journal} {\bibinfo
  {journal} {Phys. Lett.}\ }\textbf {\bibinfo {volume} {B696}},\ \bibinfo
  {pages} {74} (\bibinfo {year} {2011})},\ \Eprint
  {http://arxiv.org/abs/1007.4345} {arXiv:1007.4345 [hep-ph]} \BibitemShut
  {NoStop}%
%%CITATION = ARXIV:1007.4345;%%
\bibitem [{\citenamefont {Del~Nobile}\ \emph {et~al.}(2012)\citenamefont
  {Del~Nobile}, \citenamefont {Kouvaris}, \citenamefont {Panci}, \citenamefont
  {Sannino},\ and\ \citenamefont {Virkajarvi}}]{DelNobile:2012tx}%
  \BibitemOpen
  \bibfield  {author} {\bibinfo {author} {\bibfnamefont {E.}~\bibnamefont
  {Del~Nobile}}, \bibinfo {author} {\bibfnamefont {C.}~\bibnamefont
  {Kouvaris}}, \bibinfo {author} {\bibfnamefont {P.}~\bibnamefont {Panci}},
  \bibinfo {author} {\bibfnamefont {F.}~\bibnamefont {Sannino}}, \ and\
  \bibinfo {author} {\bibfnamefont {J.}~\bibnamefont {Virkajarvi}},\ }\href
  {\doibase 10.1088/1475-7516/2012/08/010} {\bibfield  {journal} {\bibinfo
  {journal} {JCAP}\ }\textbf {\bibinfo {volume} {1208}},\ \bibinfo {pages}
  {010} (\bibinfo {year} {2012})},\ \Eprint {http://arxiv.org/abs/1203.6652}
  {arXiv:1203.6652 [hep-ph]} \BibitemShut {NoStop}%
%%CITATION = ARXIV:1203.6652;%%
\bibitem [{\citenamefont {Kavanagh}\ and\ \citenamefont
  {Edwards}(2018)}]{WIMpy-code}%
  \BibitemOpen
  \bibfield  {author} {\bibinfo {author} {\bibfnamefont {B.~J.}\ \bibnamefont
  {Kavanagh}}\ and\ \bibinfo {author} {\bibfnamefont {T.~D.~P.}\ \bibnamefont
  {Edwards}},\ }\href@noop {} {\enquote {\bibinfo {title}
  {\textnormal{WIMpy\_NREFT v1.0 [Computer Software]},
  \href{https://doi.org/10.5281/zenodo.1230503}{\textnormal{doi:10.5281/zenodo.1230503}}\textnormal{.
  Available at }\url{https://github.com/bradkav/WIMpy_NREFT}},}\ } (\bibinfo
  {year} {2018})\BibitemShut {NoStop}%
\bibitem [{\citenamefont {Catena}\ and\ \citenamefont
  {Schwabe}(2015)}]{Catena:2015uha}%
  \BibitemOpen
  \bibfield  {author} {\bibinfo {author} {\bibfnamefont {R.}~\bibnamefont
  {Catena}}\ and\ \bibinfo {author} {\bibfnamefont {B.}~\bibnamefont
  {Schwabe}},\ }\href {\doibase 10.1088/1475-7516/2015/04/042} {\bibfield
  {journal} {\bibinfo  {journal} {JCAP}\ }\textbf {\bibinfo {volume} {1504}},\
  \bibinfo {pages} {042} (\bibinfo {year} {2015})},\ \Eprint
  {http://arxiv.org/abs/1501.03729} {arXiv:1501.03729 [hep-ph]} \BibitemShut
  {NoStop}%
%%CITATION = ARXIV:1501.03729;%%
\bibitem [{\citenamefont {Read}(2014)}]{Read:2014qva}%
  \BibitemOpen
  \bibfield  {author} {\bibinfo {author} {\bibfnamefont {J.~I.}\ \bibnamefont
  {Read}},\ }\href {\doibase 10.1088/0954-3899/41/6/063101} {\bibfield
  {journal} {\bibinfo  {journal} {J. Phys.}\ }\textbf {\bibinfo {volume}
  {G41}},\ \bibinfo {pages} {063101} (\bibinfo {year} {2014})},\ \Eprint
  {http://arxiv.org/abs/1404.1938} {arXiv:1404.1938 [astro-ph.GA]} \BibitemShut
  {NoStop}%
%%CITATION = ARXIV:1404.1938;%%
\bibitem [{\citenamefont {Green}(2017)}]{Green:2017odb}%
  \BibitemOpen
  \bibfield  {author} {\bibinfo {author} {\bibfnamefont {A.~M.}\ \bibnamefont
  {Green}},\ }\href {\doibase 10.1088/1361-6471/aa7819} {\bibfield  {journal}
  {\bibinfo  {journal} {J. Phys.}\ }\textbf {\bibinfo {volume} {G44}},\
  \bibinfo {pages} {084001} (\bibinfo {year} {2017})},\ \Eprint
  {http://arxiv.org/abs/1703.10102} {arXiv:1703.10102 [astro-ph.CO]}
  \BibitemShut {NoStop}%
%%CITATION = ARXIV:1703.10102;%%
\bibitem [{\citenamefont {Akerib}\ \emph {et~al.}(2015)\citenamefont {Akerib}
  \emph {et~al.}}]{Akerib:2015cja}%
  \BibitemOpen
  \bibfield  {author} {\bibinfo {author} {\bibfnamefont {D.~S.}\ \bibnamefont
  {Akerib}} \emph {et~al.} (\bibinfo {collaboration} {LZ}),\ }\href@noop {} {\
  (\bibinfo {year} {2015})},\ \Eprint {http://arxiv.org/abs/1509.02910}
  {arXiv:1509.02910 [physics.ins-det]} \BibitemShut {NoStop}%
%%CITATION = ARXIV:1509.02910;%%
\bibitem [{\citenamefont {Aalbers}\ \emph {et~al.}(2016)\citenamefont {Aalbers}
  \emph {et~al.}}]{Aalbers:2016jon}%
  \BibitemOpen
  \bibfield  {author} {\bibinfo {author} {\bibfnamefont {J.}~\bibnamefont
  {Aalbers}} \emph {et~al.} (\bibinfo {collaboration} {DARWIN}),\ }\href
  {\doibase 10.1088/1475-7516/2016/11/017} {\bibfield  {journal} {\bibinfo
  {journal} {JCAP}\ }\textbf {\bibinfo {volume} {1611}},\ \bibinfo {pages}
  {017} (\bibinfo {year} {2016})},\ \Eprint {http://arxiv.org/abs/1606.07001}
  {arXiv:1606.07001 [astro-ph.IM]} \BibitemShut {NoStop}%
%%CITATION = ARXIV:1606.07001;%%
\bibitem [{\citenamefont {Plante}(2016)}]{XENONnT}%
  \BibitemOpen
  \bibfield  {author} {\bibinfo {author} {\bibfnamefont {G.}~\bibnamefont
  {Plante}} (\bibinfo {collaboration} {XENON}),\ }\href@noop {} {}\bibinfo
  {howpublished} {\url{https://conferences.pa.ucla.edu/dm16/talks/plante.pdf}}
  (\bibinfo {year} {2016})\BibitemShut {NoStop}%
\bibitem [{\citenamefont {Akerib}\ \emph {et~al.}(2018)\citenamefont {Akerib}
  \emph {et~al.}}]{Akerib:2018lyp}%
  \BibitemOpen
  \bibfield  {author} {\bibinfo {author} {\bibfnamefont {D.~S.}\ \bibnamefont
  {Akerib}} \emph {et~al.} (\bibinfo {collaboration} {LUX-ZEPLIN}),\
  }\href@noop {} {\  (\bibinfo {year} {2018})},\ \Eprint
  {http://arxiv.org/abs/1802.06039} {arXiv:1802.06039 [astro-ph.IM]}
  \BibitemShut {NoStop}%
%%CITATION = ARXIV:1802.06039;%%
\bibitem [{\citenamefont {Aalseth}\ \emph {et~al.}(2018)\citenamefont {Aalseth}
  \emph {et~al.}}]{Aalseth:2017fik}%
  \BibitemOpen
  \bibfield  {author} {\bibinfo {author} {\bibfnamefont {C.~E.}\ \bibnamefont
  {Aalseth}} \emph {et~al.},\ }\href {\doibase 10.1140/epjp/i2018-11973-4}
  {\bibfield  {journal} {\bibinfo  {journal} {Eur. Phys. J. Plus}\ }\textbf
  {\bibinfo {volume} {133}},\ \bibinfo {pages} {131} (\bibinfo {year}
  {2018})},\ \Eprint {http://arxiv.org/abs/1707.08145} {arXiv:1707.08145
  [physics.ins-det]} \BibitemShut {NoStop}%
%%CITATION = ARXIV:1707.08145;%%
\bibitem [{\citenamefont {Lewin}\ and\ \citenamefont
  {Smith}(1996)}]{Lewin:1995rx}%
  \BibitemOpen
  \bibfield  {author} {\bibinfo {author} {\bibfnamefont {J.~D.}\ \bibnamefont
  {Lewin}}\ and\ \bibinfo {author} {\bibfnamefont {P.~F.}\ \bibnamefont
  {Smith}},\ }\href {\doibase 10.1016/S0927-6505(96)00047-3} {\bibfield
  {journal} {\bibinfo  {journal} {Astropart. Phys.}\ }\textbf {\bibinfo
  {volume} {6}},\ \bibinfo {pages} {87} (\bibinfo {year} {1996})}\BibitemShut
  {NoStop}%
%%CITATION = APHYE,6,87;%%
\bibitem [{\citenamefont {Green}(2007)}]{Green:2007rb}%
  \BibitemOpen
  \bibfield  {author} {\bibinfo {author} {\bibfnamefont {A.~M.}\ \bibnamefont
  {Green}},\ }\href {\doibase 10.1088/1475-7516/2007/08/022} {\bibfield
  {journal} {\bibinfo  {journal} {JCAP}\ }\textbf {\bibinfo {volume} {0708}},\
  \bibinfo {pages} {022} (\bibinfo {year} {2007})},\ \Eprint
  {http://arxiv.org/abs/hep-ph/0703217} {arXiv:hep-ph/0703217 [hep-ph]}
  \BibitemShut {NoStop}%
%%CITATION = HEP-PH/0703217;%%
\bibitem [{\citenamefont {Green}(2008)}]{Green:2008rd}%
  \BibitemOpen
  \bibfield  {author} {\bibinfo {author} {\bibfnamefont {A.~M.}\ \bibnamefont
  {Green}},\ }\href {\doibase 10.1088/1475-7516/2008/07/005} {\bibfield
  {journal} {\bibinfo  {journal} {JCAP}\ }\textbf {\bibinfo {volume} {0807}},\
  \bibinfo {pages} {005} (\bibinfo {year} {2008})},\ \Eprint
  {http://arxiv.org/abs/0805.1704} {arXiv:0805.1704 [hep-ph]} \BibitemShut
  {NoStop}%
%%CITATION = ARXIV:0805.1704;%%
\bibitem [{\citenamefont {Newstead}\ \emph {et~al.}(2013)\citenamefont
  {Newstead}, \citenamefont {Jacques}, \citenamefont {Krauss}, \citenamefont
  {Dent},\ and\ \citenamefont {Ferrer}}]{Newstead:2013pea}%
  \BibitemOpen
  \bibfield  {author} {\bibinfo {author} {\bibfnamefont {J.~L.}\ \bibnamefont
  {Newstead}}, \bibinfo {author} {\bibfnamefont {T.~D.}\ \bibnamefont
  {Jacques}}, \bibinfo {author} {\bibfnamefont {L.~M.}\ \bibnamefont {Krauss}},
  \bibinfo {author} {\bibfnamefont {J.~B.}\ \bibnamefont {Dent}}, \ and\
  \bibinfo {author} {\bibfnamefont {F.}~\bibnamefont {Ferrer}},\ }\href
  {\doibase 10.1103/PhysRevD.88.076011} {\bibfield  {journal} {\bibinfo
  {journal} {Phys. Rev.}\ }\textbf {\bibinfo {volume} {D88}},\ \bibinfo {pages}
  {076011} (\bibinfo {year} {2013})},\ \Eprint {http://arxiv.org/abs/1306.3244}
  {arXiv:1306.3244 [astro-ph.CO]} \BibitemShut {NoStop}%
%%CITATION = ARXIV:1306.3244;%%
\bibitem [{\citenamefont {Oliphant}(2015)}]{Oliphant:2015:GN:2886196}%
  \BibitemOpen
  \bibfield  {author} {\bibinfo {author} {\bibfnamefont {T.~E.}\ \bibnamefont
  {Oliphant}},\ }\href@noop {} {\emph {\bibinfo {title} {Guide to NumPy}}},\
  \bibinfo {edition} {2nd}\ ed.\ (\bibinfo  {publisher} {CreateSpace
  Independent Publishing Platform},\ \bibinfo {address} {USA},\ \bibinfo {year}
  {2015})\BibitemShut {NoStop}%
\bibitem [{\citenamefont {Pedregosa}\ \emph {et~al.}(2011)\citenamefont
  {Pedregosa}, \citenamefont {Varoquaux}, \citenamefont {Gramfort},
  \citenamefont {Michel}, \citenamefont {Thirion}, \citenamefont {Grisel},
  \citenamefont {Blondel}, \citenamefont {Prettenhofer}, \citenamefont {Weiss},
  \citenamefont {Dubourg}, \citenamefont {Vanderplas}, \citenamefont {Passos},
  \citenamefont {Cournapeau}, \citenamefont {Brucher}, \citenamefont {Perrot},\
  and\ \citenamefont {Duchesnay}}]{scikit-learn}%
  \BibitemOpen
  \bibfield  {author} {\bibinfo {author} {\bibfnamefont {F.}~\bibnamefont
  {Pedregosa}}, \bibinfo {author} {\bibfnamefont {G.}~\bibnamefont
  {Varoquaux}}, \bibinfo {author} {\bibfnamefont {A.}~\bibnamefont {Gramfort}},
  \bibinfo {author} {\bibfnamefont {V.}~\bibnamefont {Michel}}, \bibinfo
  {author} {\bibfnamefont {B.}~\bibnamefont {Thirion}}, \bibinfo {author}
  {\bibfnamefont {O.}~\bibnamefont {Grisel}}, \bibinfo {author} {\bibfnamefont
  {M.}~\bibnamefont {Blondel}}, \bibinfo {author} {\bibfnamefont
  {P.}~\bibnamefont {Prettenhofer}}, \bibinfo {author} {\bibfnamefont
  {R.}~\bibnamefont {Weiss}}, \bibinfo {author} {\bibfnamefont
  {V.}~\bibnamefont {Dubourg}}, \bibinfo {author} {\bibfnamefont
  {J.}~\bibnamefont {Vanderplas}}, \bibinfo {author} {\bibfnamefont
  {A.}~\bibnamefont {Passos}}, \bibinfo {author} {\bibfnamefont
  {D.}~\bibnamefont {Cournapeau}}, \bibinfo {author} {\bibfnamefont
  {M.}~\bibnamefont {Brucher}}, \bibinfo {author} {\bibfnamefont
  {M.}~\bibnamefont {Perrot}}, \ and\ \bibinfo {author} {\bibfnamefont
  {E.}~\bibnamefont {Duchesnay}},\ }\href@noop {} {\bibfield  {journal}
  {\bibinfo  {journal} {Journal of Machine Learning Research}\ }\textbf
  {\bibinfo {volume} {12}},\ \bibinfo {pages} {2825} (\bibinfo {year}
  {2011})}\BibitemShut {NoStop}%
\bibitem [{\citenamefont {Weisstein}()}]{weissteinhypersphere}%
  \BibitemOpen
  \bibfield  {author} {\bibinfo {author} {\bibfnamefont {E.~W.}\ \bibnamefont
  {Weisstein}},\ }\href {http://mathworld.wolfram.com/HyperspherePacking.html}
  {\enquote {\bibinfo {title} {Hypersphere packing},}\ }\BibitemShut {NoStop}%
\bibitem [{\citenamefont {Bovy}\ \emph {et~al.}(2012)\citenamefont {Bovy} \emph
  {et~al.}}]{Bovy:2012ba}%
  \BibitemOpen
  \bibfield  {author} {\bibinfo {author} {\bibfnamefont {J.}~\bibnamefont
  {Bovy}} \emph {et~al.},\ }\href {\doibase 10.1088/0004-637X/759/2/131}
  {\bibfield  {journal} {\bibinfo  {journal} {Astrophys. J.}\ }\textbf
  {\bibinfo {volume} {759}},\ \bibinfo {pages} {131} (\bibinfo {year}
  {2012})},\ \Eprint {http://arxiv.org/abs/1209.0759} {arXiv:1209.0759
  [astro-ph.GA]} \BibitemShut {NoStop}%
%%CITATION = ARXIV:1209.0759;%%
\bibitem [{\citenamefont {Koposov}\ \emph {et~al.}(2010)\citenamefont
  {Koposov}, \citenamefont {Rix},\ and\ \citenamefont {Hogg}}]{Koposov:2009hn}%
  \BibitemOpen
  \bibfield  {author} {\bibinfo {author} {\bibfnamefont {S.~E.}\ \bibnamefont
  {Koposov}}, \bibinfo {author} {\bibfnamefont {H.-W.}\ \bibnamefont {Rix}}, \
  and\ \bibinfo {author} {\bibfnamefont {D.~W.}\ \bibnamefont {Hogg}},\ }\href
  {\doibase 10.1088/0004-637X/712/1/260} {\bibfield  {journal} {\bibinfo
  {journal} {Astrophys. J.}\ }\textbf {\bibinfo {volume} {712}},\ \bibinfo
  {pages} {260} (\bibinfo {year} {2010})},\ \Eprint
  {http://arxiv.org/abs/0907.1085} {arXiv:0907.1085 [astro-ph.GA]} \BibitemShut
  {NoStop}%
%%CITATION = ARXIV:0907.1085;%%
\bibitem [{\citenamefont {Piffl}\ \emph {et~al.}(2014)\citenamefont {Piffl}
  \emph {et~al.}}]{Piffl:2013mla}%
  \BibitemOpen
  \bibfield  {author} {\bibinfo {author} {\bibfnamefont {T.}~\bibnamefont
  {Piffl}} \emph {et~al.},\ }\href {\doibase 10.1051/0004-6361/201322531}
  {\bibfield  {journal} {\bibinfo  {journal} {Astron. Astrophys.}\ }\textbf
  {\bibinfo {volume} {562}},\ \bibinfo {pages} {A91} (\bibinfo {year}
  {2014})},\ \Eprint {http://arxiv.org/abs/1309.4293} {arXiv:1309.4293
  [astro-ph.GA]} \BibitemShut {NoStop}%
%%CITATION = ARXIV:1309.4293;%%
\bibitem [{\citenamefont {Peter}(2010)}]{Peter:2009ak}%
  \BibitemOpen
  \bibfield  {author} {\bibinfo {author} {\bibfnamefont {A.~H.~G.}\
  \bibnamefont {Peter}},\ }\href {\doibase 10.1103/PhysRevD.81.087301}
  {\bibfield  {journal} {\bibinfo  {journal} {Phys. Rev.}\ }\textbf {\bibinfo
  {volume} {D81}},\ \bibinfo {pages} {087301} (\bibinfo {year} {2010})},\
  \Eprint {http://arxiv.org/abs/0910.4765} {arXiv:0910.4765 [astro-ph.CO]}
  \BibitemShut {NoStop}%
%%CITATION = ARXIV:0910.4765;%%
\bibitem [{\citenamefont {Strigari}\ and\ \citenamefont
  {Trotta}(2009)}]{Strigari:2009zb}%
  \BibitemOpen
  \bibfield  {author} {\bibinfo {author} {\bibfnamefont {L.~E.}\ \bibnamefont
  {Strigari}}\ and\ \bibinfo {author} {\bibfnamefont {R.}~\bibnamefont
  {Trotta}},\ }\href {\doibase 10.1088/1475-7516/2009/11/019} {\bibfield
  {journal} {\bibinfo  {journal} {JCAP}\ }\textbf {\bibinfo {volume} {0911}},\
  \bibinfo {pages} {019} (\bibinfo {year} {2009})},\ \Eprint
  {http://arxiv.org/abs/0906.5361} {arXiv:0906.5361 [astro-ph.HE]} \BibitemShut
  {NoStop}%
%%CITATION = ARXIV:0906.5361;%%
\bibitem [{\citenamefont {Pato}\ \emph {et~al.}(2013)\citenamefont {Pato},
  \citenamefont {Strigari}, \citenamefont {Trotta},\ and\ \citenamefont
  {Bertone}}]{Pato:2012fw}%
  \BibitemOpen
  \bibfield  {author} {\bibinfo {author} {\bibfnamefont {M.}~\bibnamefont
  {Pato}}, \bibinfo {author} {\bibfnamefont {L.~E.}\ \bibnamefont {Strigari}},
  \bibinfo {author} {\bibfnamefont {R.}~\bibnamefont {Trotta}}, \ and\ \bibinfo
  {author} {\bibfnamefont {G.}~\bibnamefont {Bertone}},\ }\href {\doibase
  10.1088/1475-7516/2013/02/041} {\bibfield  {journal} {\bibinfo  {journal}
  {JCAP}\ }\textbf {\bibinfo {volume} {1302}},\ \bibinfo {pages} {041}
  (\bibinfo {year} {2013})},\ \Eprint {http://arxiv.org/abs/1211.7063}
  {arXiv:1211.7063 [astro-ph.CO]} \BibitemShut {NoStop}%
%%CITATION = ARXIV:1211.7063;%%
\bibitem [{\citenamefont {Schumann}\ \emph {et~al.}(2015)\citenamefont
  {Schumann}, \citenamefont {Baudis}, \citenamefont {Bütikofer}, \citenamefont
  {Kish},\ and\ \citenamefont {Selvi}}]{Schumann:2015cpa}%
  \BibitemOpen
  \bibfield  {author} {\bibinfo {author} {\bibfnamefont {M.}~\bibnamefont
  {Schumann}}, \bibinfo {author} {\bibfnamefont {L.}~\bibnamefont {Baudis}},
  \bibinfo {author} {\bibfnamefont {L.}~\bibnamefont {Bütikofer}}, \bibinfo
  {author} {\bibfnamefont {A.}~\bibnamefont {Kish}}, \ and\ \bibinfo {author}
  {\bibfnamefont {M.}~\bibnamefont {Selvi}},\ }\href {\doibase
  10.1088/1475-7516/2015/10/016} {\bibfield  {journal} {\bibinfo  {journal}
  {JCAP}\ }\textbf {\bibinfo {volume} {1510}},\ \bibinfo {pages} {016}
  (\bibinfo {year} {2015})},\ \Eprint {http://arxiv.org/abs/1506.08309}
  {arXiv:1506.08309 [physics.ins-det]} \BibitemShut {NoStop}%
%%CITATION = ARXIV:1506.08309;%%
\bibitem [{\citenamefont {Feng}\ \emph {et~al.}(2011)\citenamefont {Feng},
  \citenamefont {Kumar}, \citenamefont {Marfatia},\ and\ \citenamefont
  {Sanford}}]{Feng:2011vu}%
  \BibitemOpen
  \bibfield  {author} {\bibinfo {author} {\bibfnamefont {J.~L.}\ \bibnamefont
  {Feng}}, \bibinfo {author} {\bibfnamefont {J.}~\bibnamefont {Kumar}},
  \bibinfo {author} {\bibfnamefont {D.}~\bibnamefont {Marfatia}}, \ and\
  \bibinfo {author} {\bibfnamefont {D.}~\bibnamefont {Sanford}},\ }\href
  {\doibase 10.1016/j.physletb.2011.07.083} {\bibfield  {journal} {\bibinfo
  {journal} {Phys. Lett.}\ }\textbf {\bibinfo {volume} {B703}},\ \bibinfo
  {pages} {124} (\bibinfo {year} {2011})},\ \Eprint
  {http://arxiv.org/abs/1102.4331} {arXiv:1102.4331 [hep-ph]} \BibitemShut
  {NoStop}%
%%CITATION = ARXIV:1102.4331;%%
\bibitem [{\citenamefont {Agnes}\ \emph {et~al.}(2015)\citenamefont {Agnes}
  \emph {et~al.}}]{Agnes:2014bvk}%
  \BibitemOpen
  \bibfield  {author} {\bibinfo {author} {\bibfnamefont {P.}~\bibnamefont
  {Agnes}} \emph {et~al.} (\bibinfo {collaboration} {DarkSide}),\ }\href
  {\doibase 10.1016/j.physletb.2015.03.012} {\bibfield  {journal} {\bibinfo
  {journal} {Phys. Lett.}\ }\textbf {\bibinfo {volume} {B743}},\ \bibinfo
  {pages} {456} (\bibinfo {year} {2015})},\ \Eprint
  {http://arxiv.org/abs/1410.0653} {arXiv:1410.0653 [astro-ph.CO]} \BibitemShut
  {NoStop}%
%%CITATION = ARXIV:1410.0653;%%
\end{thebibliography}%

\appendix

\section*{Supplemental Material}

\section{Technical details}
\label{sec:apx1}

First, we generate a large number of points in the parameter space of model
$\M$, $\vec\theta_i \in \Omega_\M$ (typically of the order $10^5$).  The
details of the distribution of the points do not matter in the limit of a large
number of parameter points.  Each point is projected onto the corresponding
euclideanized signal, $\vec x_i = \vec x(\vec \theta_i)$.  This projection
depends on the details of the detector.  If multiple experiments are used, the
corresponding vectors are concatenated.  Euclideanized signals are generated
using \texttt{swordfish}~\cite{Edwards:2017kqw} (see paper for technical
details).  This process essentially provides a sample of the model parameter
space $\M$ embedded in the (usually higher dimensional) euclideanized signal
space. In addition, the mapping between these spaces is known if $\M$ is sufficiently sampled. This sample of parameter points and corresponding euclideanized signals are the basis for the various estimation techniques used in this work.

\medskip

\textit{Confidence contours.} Given the sample of projected points, $\vec x_i$,
it is now straightforward to generate expected contours around any parameter
$\vec\theta_0\in\Omega_\M$ in the model parameter space.  Such regions are for
instance shown in Fig.~\ref{fig:regions}. To this end, we first calculate the
projected signal $\vec x_0 = \vec x(\vec\theta_0)$.  Then, using standard
nearest neighbor finder algorithms \cite{scikit-learn}, we identify the set of points
$\vec\theta_i$ that are within a radius $r_{\alpha}(\M)$ of point $\vec\theta_0$.
Here, $r_{\alpha}(\M)$ depends on the dimensionality of the parameter space $\Omega_\M$
as well as the significance level of interest, $r_{\alpha}(\M) = \sqrt{\chi^2_{k=d,\mathrm{ISF}}(1-\alpha)}$ where $\chi^2_{k=d,\mathrm{ISF}}$ is the inverse survival function of the Chi-squared distribution with $k=d$ degrees of freedom. For the 3-dim models that we
consider in this paper, and a significance level of $\alpha = 0.045$, we have
for instance $r_{\alpha}(\M) = 2.84$.  Now, the points in the model parameter space $\Omega_\M$ that belong
to the confidence region can be simply identified by back-projecting the
nearest neighbors in euclideanized signal space (obviously this back-projecting
just requires a look-up in the original list of model parameters).  In this
way, the generation of confidence regions around arbitrary signal points is
efficiently achieved.

\textit{Number of distinct signals.} For Fig.~\ref{fig:Venn} we are interested
in the (maximum) number of points that can populate the model parameter space
$\Omega_\M$ such that the model points can be discriminated in the sense of
Eq.~\eqref{eqn:TS}.  This is equivalent to finding the maximum number of points
in the euclideanized signal space that can populate the embedding of
$\Omega_\M$ such that their mutual distance is at least $r_{\alpha}(\M)$.  
We derive an estimate for this number with the following procedure: For each
projected sample point $\vec x_i$, we calculate the number of nearest neighbors $N_i$
within a distance $r_{\alpha}(\M)/2$.  This can be rather efficiently done using
standard clustering algorithms.  Now, we assign a weight to point $\vec x_i$,
which is simply given by $w_i = 1/N_i$.  It hence corresponds to the
`fractional contribution' of a single parameter point to a confidence region.
The number of distinguishable signals is given by the sum
\begin{equation}
  \nu_{\M, X}^\alpha(\Omega_\M) = c_{ff}\sum_i w_i\;.
  \label{eqn:sum}
\end{equation}
Here, $c_{ff}$ is a filling factor correction related to the packing density of
hyperspheres in a $d$-dim parameter space.  For the 3-dim models in which we
are often interested, this number is given by $c_{ff} = 0.74$ \cite{weissteinhypersphere}.  We find that this prescription provides an efficient and
reliable way to estimate the number of distinct signals of a model.  The main
requirement in the calculation is that each of the potential confidence regions
contains a sufficiently large (typically at least ten) number of samples.
Adding more points to the original list would then not affect the result
anymore. We tested the stability of our results by doubling the number of sampled points in various examples from the text. The results remained unchanged in the limit of ten points per distinct region.

\medskip

\textit{Distinct signals compatible with $\mathcal{H}_0$}. We are interested in
the fraction of the distinct signal points that are consistent with a
null hypothesis that is defined as a lower dimensional subspace of the full
model parameter space, $\Omega_\S \in \Omega_\M$.  Here, we call a point in
$\Omega_\M$ `consistent' with $\Omega_\S$ if the composite null hypothesis
$\Omega_\S$ cannot be excluded against the alternative hypothesis $\Omega_\M$.
To estimate this number, we first generate a large number of points in
$\Omega_\S$.  We then collect all points from the original sample of $\Omega_M$
whose minimum distance to any of the points from $\Omega_\S$ is smaller than
the threshold values $r_{\alpha}(\M, \S)$.  Here, the threshold is derived from a
$\chi^2_k$ distribution with $k$ degrees of freedom, where $k$ is the
difference in the dimensionality of $\Omega_\M$ and $\Omega_S$ (for the
examples in the paper, we usually have $k=1$, and hence $r_{\alpha}(\M, \S) = 2$).  The number of distinct signals that are
compatible with the null hypothesis $\Omega_\S$ is then simply obtained by
restricting the sum in Eq.~\eqref{eqn:sum} to the points within the shell around $\Omega_\S$.

\medskip

\textit{Parameter ranges and nuisance parameters.} Finally, the contours in
Fig.~\ref{fig:comp} and Fig.~\ref{fig:mass} are generated by identifying all
points that are consistent with the indicated null hypotheses, as described in the
previous paragraph.  However, in these figures we also take into account the
effects of DM halo uncertainties, as described in the main text (this is not
easily possible when calculating the number of dimensions).  To this end, we generate for each
point $\vec x_i \in \Omega_\M$ several euclideanized signals corresponding to
various randomly selected DM halo configurations.  Again, the specific
distribution of these points does not matter as long as the number is large
enough to sufficiently cover the various halo configurations.  In order to
incorporate external constraints on the DM halo parameters, we add an
additional contribution to the euclideanized distance calculation, which is
just given by $(\eta_i - \bar{\eta})^2/\sigma_\eta^2$, where $\bar\eta$ and
$\sigma_\eta^2$ are the mean and variance of the nuisance parameter, and $\eta_i$ is
the value of the nuisance parameter for a specific point $i$. The contribution to the $\vec{x}(\vec{\theta})$ is a simple concatenation of $(\eta_i - \bar{\eta})/\sigma_\eta$ with the Euclideanised signal.

For a large number of sampled points this approach exactly matches a profile log-likelihood
analysis.  We check this limit is saturated by increasing the
number of sampled points until our results do not noticeably change.

\section{Dark matter signal modeling}
\label{sec:apx2}

\textit{Halo Uncertainties.} We incorporate halo uncertainties \cite{Green:2017odb} by assuming Gaussian
likelihood distributions for three parameters of the Maxwellian velocity
distribution of DM: the Sun's speed $v_\odot= (242 \pm 10)\, \mathrm{km/s}$
\cite{Bovy:2012ba}, the local circular speed $v_\mathrm{c} = (220 \pm 18) \,
\mathrm{km/s}$  \cite{Koposov:2009hn}, and the Galactic escape speed
$v_\mathrm{esc} =(533 \pm 54)\, \mathrm{km/s}$ \cite{Piffl:2013mla}. We assume
that these uncertainties are uncorrelated \cite{Peter:2009ak}, though in general correlations coming from the
modeling of the Milky Way halo can be included
\cite{Strigari:2009zb,Pato:2012fw}. We sample from these distributions as nuisance parameters
in our signal calculation and include an additional penalization term to the euclideanized signal in Eq.~\eqref{eqn:TS}.

\medskip

\textit{Detector specifications.} We implement a simplified XENON1T for which we adopt
an S1-only analysis, full details of which are given in Sec.~IIIB of \cite{Edwards:2017kqw}. For the recoil spectrum $\mathrm{d}R/\mathrm{d}S_1$, we use 19 bins linearly
spaced between 3 and 70 PE (corresponding to nuclear recoil energies $E_R \in [5,\, 40] \,\mathrm{keV}$). The number of bins was chosen for computational efficiency with no
noticeable loss in accuracy. We have checked that including a 20\% energy resolution \cite{Schumann:2015cpa} and increasing the number of bins should have no substantial effect on our results. Background distributions as a function of $S_1$ are described in Fig. 3 of \cite{Aprile:2017iyp} for
which we assume 1\% uncertainty on all components separately. We also sum over different Xenon isotopes, weighting by their naturally-occurring
mass fractions \cite{Feng:2011vu}.

For our future Xenon detector we scale up the observation time of XENON1T-2017 by a factor of
100, assuming that background rates stay constant. This exposure
roughly corresponds to that expected for the full run of the XENONnT experiment
\cite{XENONnT}, so we will refer to this future detector as \textit{XENONnT}.

\textit{DarkSide20k:} We directly use the recoil energy spectrum $\mathrm{d}R/\mathrm{d}E_R$ as our signal, assuming that the only
relevant isotope is Argon-40. We follow the specifications of the Darkside50 detector \cite{Agnes:2014bvk}, with the nuclear recoil efficiency taken from Fig.~6 in~\cite{Agnes:2014bvk}. The background is assumed to be flat across the entire energy range
with an expected $0.1$ events (with $10\%$ uncertainty) over $1422 \textrm{\ kg\ days}$ of observation. We assume 19 linearly spaced bins between 32 and 200
$\textrm{keV}$. 

For our future detector, we assume an exposure of $7.3\times 10^6 \mathrm{\ kg \ days}$, corresponding to a 1-year exposure with a 20-tonne detector. We assume that the background will remain at $0.1$ events in the total exposure. This detector configuration roughly resembles \textit{DarkSide20k} \cite{Aalseth:2017fik}.

\end{document}